\documentclass[twocolumn]{aastex631}

\usepackage{graphicx}
\usepackage{amsmath}
\usepackage{bm}
\usepackage{multirow}
\usepackage{float}
\makeatletter
\let\newfloat\newfloat@ltx
\makeatother
\usepackage[linesnumbered,algoruled,boxed,lined,algo2e]{algorithm2e}
\usepackage{algorithm}
\usepackage{makecell}
\usepackage{tabularx}

\hypersetup{
            colorlinks=true,
            linkcolor=blue,
            anchorcolor=blue,
            citecolor=blue}
\usepackage{hyperref}

\begin{document}

\title{Modeling the Time Evolution of Compact Binary Systems with Machine Learning}

\correspondingauthor{Jie Feng, Hong-Hao Zhang, Weipeng Lin}
\email{E-mail: fengj77@mail.sysu.edu.cn, zhh98@mail.sysu.edu.cn, linweip5@mail.sysu.edu.cn}

\affiliation{
School of Physics, Sun Yat-sen University, Guangzhou 510275, China
}%
\affiliation{
Faculty of Innovation Engineering, Macau University of Science and Technology, Avenida Wai Long, Taipa 999078, Macau}
\affil{
 School of Science, Shenzhen Campus of Sun Yat-sen University, Shenzhen 518107, China
}
\affil{
School of Physics and Astronomy, Sun Yat-sen University, Zhuhai 519000, China
}%

\author{Jianqi Yan}
\thanks{These authors contributed equally to this work.}
\affiliation{
Faculty of Innovation Engineering, Macau University of Science and Technology, Avenida Wai Long, Taipa 999078, Macau}

\author{Junjie Luo}
\thanks{These authors contributed equally to this work.}
\affiliation{
School of Physics, Sun Yat-sen University, Guangzhou 510275, China
}%

\author{Yifan Zeng}
\affil{
School of Electrical Engineering and Computer Science, Oregon State University, Corvallis, Oregon, OR 97331, USA
}%

\author{Alex P. Leung}
\affil{
Department of Physics, The University of Hong Kong, 999077, Hong Kong
}

\author[0000-0003-3649-2476]{Jie Feng}
\affil{
 School of Science, Shenzhen Campus of Sun Yat-sen University, Shenzhen 518107, China
}

\author{Hong-Hao Zhang}
\affil{
School of Physics, Sun Yat-sen University, Guangzhou 510275, China
}

\author{Weipeng Lin}
\affil{
School of Physics and Astronomy, Sun Yat-sen University, Zhuhai 519000, China
}%



\begin{abstract}

This work introduces advanced computational techniques for modeling the time evolution of compact binary systems using machine learning.
The dynamics of compact binary systems, such as black holes and neutron stars, present significant nonlinear challenges due to the strong gravitational interactions and the requirement for precise numerical simulations.
Traditional methods, like the post-Newtonian approximation, often require significant computational resources and face challenges in accuracy and efficiency.
Here, we employed machine learning algorithms, including deep learning models like Long Short-Term Memory (LSTM) and Temporal Convolutional Network (TCN), to predict the future evolution of these systems based on extensive simulation data.
Our results demonstrate that employing both LSTM and TCN even as black-box predictors for sequence prediction can also significantly improve the prediction accuracy without PINNs as PDE solvers with prior knowledge or inductive bias.
By employing LSTM and TCN, we obtained $R^2$ values of 99.74\% and 99.19\% for the evolutionary orbits of compact binaries dataset, respectively.
Our models demonstrate the ability to effectively capture the dynamics of the binaries, achieving high prediction performance with significantly reduced computational overhead by a factor of 40, compared to conventional numerical methods.
This study paves the way for more effective and computationally scalable approaches to the understanding of gravitational phenomena and predictive modeling in gravitational-wave astronomy.

\end{abstract}



\section{\label{sec1}Introduction}

The phenomenology associated with two orbiting compact objects in general relativity is indeed rich and complex, encompassing a wide range of astrophysical phenomena and mathematical challenges. At the heart of this theory lies Einstein's general theory of relativity, which not only revolutionized our understanding of gravity but also predicted the existence of gravitational waves. These waves, ripples in the fabric of spacetime, are emitted when massive objects accelerate or orbit each other. Their existence was a crucial prediction of Einstein's theory, one that eluded detection for decades until the advent of advanced interferometric detectors such as LIGO \citep{LIGOScientific:2016aoc}. The confirmation of gravitational waves in 2016 by the LIGO Scientific Collaboration and Virgo Collaboration not only validated Einstein's theory but also opened a new window to observe the universe. The motion of strong gravitational systems is difficult to describe analytically. Exact solutions to Einstein's equations in such scenarios are pretty difficult to obtain, However, there are some efforts to address this issue. For example, \cite{Wu_2022} work focused on developing these algorithms specifically for curved spacetimes, including those that describe black holes. This innovation has the potential to significantly improve the accuracy and efficiency of numerical simulations involving a strong gravitational field.

In addition, the study of compact binary systems in the context of general relativity often requires high-precision methods to accurately model their dynamics. One such approach is the post-Newtonian approximation, which provides a powerful framework for describing the motion of these systems in the weak-field and slow-motion limits (\cite{Blanchet_2006}).

Within the post-Newtonian Hamiltonian formalism, implicit algorithms (\cite{Seyrich:2013PRD, Zhong_2010}) are an optional tool, but they suffer from computational inefficiency over multiple iterations and convergence issues that make numerical solutions unavailable. As an alternative, a more efficient and stable integrator, the extended phase-space method has emerged as a valuable tool. This method, as exemplified by some recent works (\cite{Pihajoki_2014, Li_2017, Li_2019}), aims to transform the Hamiltonian into a form that facilitates the use of explicit numerical integrators like the leapfrog algorithm. In addition, \cite{Pihajoki_2014} introduced a permutation map for the momenta helps to suppress unwanted interactions between the original and extended variables.

Nevertheless, challenges arise when applying these algorithms to chaotic orbits of celestial systems. The interactions between the original and extended variables can become pronounced, leading to significant differences that violate their intended equivalence. To address this issue, midpoint and correction maps have been proposed. These maps, as exemplified by \citep{Luo:2017ApJ, Luo_2020, Luo_2022}, help to ensure the equivalence between the original and extended variables, thereby improving the accuracy and reliability of \textcolor{blue}{the extended phase-space} method.

While the Extended Phase-Space Method yields valuable numerical solutions, analyzing the vast amount of data it generates can be a challenging task. Here is where machine learning comes into play. Machine learning algorithms, such as artificial neural networks, support vector machines, and decision trees, can be employed to study the complex relationships between variables in numerical solutions. By feeding the numerical data as inputs and corresponding binary orbital parameters as outputs, machine learning algorithms can learn to recognize patterns and correlations in the data. This process allows researchers to identify significant variables influencing compact binary evolution and gain a deeper understanding of the underlying physics.

Once machine learning models are trained on the numerical solutions and the associated orbital parameters, they can be utilized to predict the future evolution of compact binaries. Given a set of initial conditions, such as masses, spins, and orbital separations, the machine learning model can forecast the binary's trajectory and parameter evolution with high accuracy. Furthermore, machine learning algorithms can explore different scenarios by varying initial conditions and studying their impact on the binary's orbit. This capability opens up exciting possibilities for understanding the influence of various physical processes on the compact binary system and provides crucial insights for gravitational wave astronomy.

Applications of deep learning and machine learning methods to solve practical problems are increasingly common in astronomy.
Liao et al. used an artificial neural network model to obtain numerically the planar periodic orbits of three-body systems with arbitrary masses \citep{liao2022three}.
\textcolor{blue}{Yan et al. employed a generative deep learning method as a data augmentation tool to increase the number of training spectrograms for glitch classification \citep{yan2022improving}.}
Additionally, an artificial neural network model dedicated to solving the chaotic three-body problem is introduced by \cite{breen2020newton}.
There are also a large number of successful implementations in deep learning time series forecasting tasks.
For instance, Park et al. proposed a method based on the Bayesian attentive neural processes for reconstructing active galactic nuclei time series to infer black hole properties \citep{park2021inferring}.
In order to implement the astronomical time series classification, \cite{jamal2020neural} discussed the performance comparison of multiple mainstream deep learning time series models on a dataset of variable stars.
These works have inspired us by showing that deep learning time series prediction can be equally applicable to predicting the future evolution of compact binaries.

By inputting time series data from compact binary systems, this study predicts future evolution using two deep learning algorithms: long short-term memory (LSTM) \citep{hochreiter1997long} and temporal convolutional network (TCN) \citep{lea2017temporal}.
This study can be considered a multivariate time series forecast, in which the predictive model is built to accelerate the process of parameter evolution and predict the future position coordinates $(q'_1, q'_2, q'_3)$ and momentum coordinates $(p'_1, p'_2, p'_3)$ based on their respective past values $(q_1, q_2, q_3)$ and $(p_1, p_2, p_3)$.
To investigate the performance of various deep network models in predicting the future evolution of compact binaries, we constructed a benchmark using the two aforementioned deep network models in this study.
We aim to present the community with a performance comparison overview of various time-series deep network models in astronomy.
Section \ref{sec31} provides further information on the use and comparison of the two deep network models.

Traditional machine learning models frequently attempt to approximate data through regression, disregarding pertinent physical constraints.
Conversely, in astrophysics, meeting fundamental physical laws and principles is crucial to tackling our target problems.
To address the issue, we drew inspiration from physics-informed neural networks (PINNs) \cite{raissi2019physics} and merged the Hamiltonian and angular momentum conservation, and canonical equations into loss function.
In our experiments, this loss function design proves to be effective in improving the prediction performance of models with deficient fitting ability.
The experimental results showed that LSTM achieved significant results in terms of both accuracy and speed.
Compared to some numerical techniques, LSTM is at least 40 times faster.
Employing machine learning techniques to enhance computational efficiency can significantly reduce the consumption of computer resources, particularly when dealing with small time steps that would otherwise lead to a marked increase in computation time.
As a result, within the same duration, we can obtain a larger volume of orbital evolution data.
In the future, our aim is to train additional models capable of inferring various parameters of the binary system—such as masses, binary separation, the direction of the gravitational wave source, and distance—directly from gravitational waveforms.
This challenging task necessitates a vast amount of orbital data, and the work presented in this article furnishes us with an efficient orbital generation tool that will forestall the prospect of prolonged wait times for orbit calculations in our future endeavors.

The structure of the remainder of this paper is as follows.
A comprehensive description and mathematical representation of the dataset is given in Section \ref{sec2}.
The introduction of our deep network architectures and the design of the loss function are presented in Section \ref{sec31} and Section \ref{sec32}, respectively.
In Section \ref{sec33}, we introduce the data pre-processing method and feature augmentation techniques.
In Section \ref{sec34}, the details of all the training experiments are presented.
Then, the experimental results are provided and discussed in Section \ref{sec4}, where TCN and LSTM can successfully predict the time evolution of compact binary systems.
Finally, in Section \ref{sec5}, we conclude our work and discuss possible future directions.

\section{\label{sec2} The symplectic-like algorithm on compact binaries and simulations}
\subsection{\label{sec21}Extended phase space method in compact binaries}

In the context of a non-spinning compact binary system without dissipated terms, the Lagrangian description at the first post-Newtonian (1PN) approximation level, as detailed in \cite{Blanchet:2002mb} is derived by incorporating relativistic corrections to the purely Newtonian orbital dynamics. Employing natural units where the fundamental constants of light speed, \( c \), and the gravitational constant, \( G \), are set to unity, \( c = G = 1 \), the system's dynamical evolution is governed by a Lagrangian that encapsulates the refined interplay between the binary components.

This Lagrangian, which serves as the foundation for formulating the equations of motion, accounts for the mutual gravitational interactions between the pair of massive objects in a manner that transcends the point-mass approximation and incorporates the leading-order general relativistic effects. In this framework, the binary constituents are treated as point masses with well-defined positions and velocities, but their individual spins are neglected to maintain the focus on purely orbital dynamics. The variables' evolution is controlled by the following Lagrangian

\begin{eqnarray}
L=L_{N}+L_{1PN},
\end{eqnarray}
where 
\begin{eqnarray}
L_{N}=\frac{\mathbf{\dot{r}}^{2}}{2}+\frac{1}{r}, 
\end{eqnarray}
\begin{eqnarray}
L_{1PN} &=& \frac{1}{8}(1-3\eta)\mathbf{\dot{r}}^{4}+\frac{1}{2}[(3+\eta)\mathbf{\dot{r}}^{2}
\nonumber \\
&& +\eta(\textbf{n}\cdot\mathbf{\dot{r}})^{2}]\frac{1}{r} -\frac{1}{2r^{2}},
\end{eqnarray}

where \( \eta = \frac{m_1 m_2}{M^2} \) and the two bodies' mass are denoted as \( m_1 \) and \( m_2 \), while total mass are \( M = m_1 + m_2 \). \( \mu = \frac{m_1 m_2}{M} \) is reduced mass, the mass ratio \( \beta = \frac{m_1}{m_2} \) and $n=\frac{\mathbf{r}}{r}$. In addition, the Lagrangian \( L \) serves as the analytical treatment of the non-spinning compact binary system's orbital dynamics within the 1PN approximation, it is often advantageous to work with an alternative yet equivalent formulation: the Hamiltonian \( H \). The Hamiltonian is derived through the Legendre transformation, 
\begin{eqnarray}
H=\mathbf{p}\cdot \mathbf{\dot{r}}-L,
\end{eqnarray}
\begin{eqnarray}
\mathbf{p}=\partial L/\partial \mathbf{\dot{r}}.
\end{eqnarray}
Then it can be deduced that the 1PN Hamiltonian \citep{Wu_2015APS},
\begin{eqnarray}
\label{2pnhami}
H=H_{N}+H_{1PN}, \label{eq:H}
\end{eqnarray}

Respectively the sub-Hamiltonians of the Equation \ref{eq:H} are expressed as
 \begin{eqnarray}
H_{N}=T(\textbf{p})+V(\textbf{r})=\frac{\textbf{p}^{2}}{2}-\frac{1}{r},
\end{eqnarray}

\begin{eqnarray}
H_{1PN} &=& \frac{1}{8}(3\eta-1)\textbf{p}^{4}-\frac{1}{2}[(3+\eta)\textbf{p}^{2} \nonumber \\
&& +\eta(\textbf{n}\cdot\textbf{p})^{2}]\frac{1}{r}+\frac{1}{2r^{2}},
\end{eqnarray}

The extended phase-space method, as introduced by \citet{Pihajoki_2014}, presents a compelling alternative approach to tackling non-separable Hamiltonian systems, which are as functions of the position vector \( \mathbf{r} \) and its conjugate momentum \( \mathbf{p} \), and unable to decompose as two or more into simpler, integrable part. Central to this methodology is the duplication and subsequent manipulation of the original pair of canonical variables \( (\mathbf{r}, \mathbf{p}) \). A copy of these variables, denoted by \( (\widetilde{\mathbf{r}}, \widetilde{\mathbf{p}}) \), is introduced, effectively doubling the dimensionality of the phase space. Then, we employ two sets of canonically conjugate variables $(\textbf{r}, \textbf{p})$ and ($\widetilde{\textbf{r}}$, $\widetilde{\textbf{p}}$) in the extended phase space to reconstruct a new Hamiltonian representation given by:
\begin{eqnarray}
       \widetilde{H}(\textbf{r},\widetilde{\textbf{r}},\textbf{p},
       \widetilde{\textbf{p}})=H_{1}(\textbf{r},
       \widetilde{\textbf{p}})+H_{2}(\widetilde{\textbf{r}},\textbf{p}).
\end{eqnarray}
Both Hamiltonians, denoted by $H_1$ and $H_2$, ought to be equivalent to the initial Hamiltonian $H$. Upon completion of the aforementioned manipulations, the resultant composite Hamiltonian $\widetilde{H}$ shall consist of two integrable constituents, thereby enabling the employment of the standard second-order leapfrog integration scheme, as detailed in \cite{Pihajoki_2014}, which can be written as:
\begin{eqnarray}
\mathbf{A}_2(h)=\textbf{H}_{2}(\frac{h}{2})\textbf{H}_{1}(h)\textbf{H}_{2}(\frac{h}{2}),
\end{eqnarray}
The Hamiltonian operators $\textbf{H}_1$ and $\textbf{H}_2$, which are functions of the time step $h$, are intended to replicate the behavior of the original Hamiltonian $\textbf{H}$. Crucially, the solution pairs $(\textbf{r}, \widetilde{\textbf{p}})$ and $(\widetilde{\textbf{r}}, \textbf{p})$ derived from $\textbf{H}_1$ and $\textbf{H}_2$, respectively, should coincide at each discrete time step. Nonetheless, despite this initial agreement, they tend to diverge significantly over time due to the intricate coupling between the solutions $(\textbf{r}, \widetilde{\textbf{p}})$ governed by $\textbf{H}_1$ and those $(\widetilde{\textbf{r}}, \textbf{p})$ evolving under $\textbf{H}_2$.

In an attempt to address this divergence issue, \cite{Pihajoki_2014} puts forth the momentum permutation map, aimed at maintaining the equivalence of the solutions from $H_1$ and $H_2$. However, this approach falls short in accurately simulating chaotic orbits of compact binaries, particularly when dealing with systems. The deficiency is rectified by the subsequent introduction of the correction map, as presented in \cite{Luo_2020},

\begin{eqnarray}
\textbf{M}_1=\left(\begin{array}{cccc}
\frac{\textbf{1}}{2},  \frac{\textbf{1}}{2},  \textbf{0},  \textbf{0} \\
\frac{\textbf{1}}{2}, \frac{\textbf{1}}{2},  \textbf{0},  \textbf{0} \\
\textbf{0},  \textbf{0}, \mathbf{\alpha}, \mathbf{\alpha} \\
\textbf{0},  \textbf{0},   \mathbf{\alpha},\mathbf{\alpha}
\end{array}\right).
\end{eqnarray}
where $\alpha$ serves as a scaling parameter, the determination of which can be accomplished either through the exploitation of a conserved quantity inherent to the system or via a bespoke methodology devised by the investigator. The ensuing numerical integration scheme, known as the extended phase-space leapfrog algorithm incorporating a correction map, can be formulated as follows:
\begin{eqnarray}
       \mathbf{C}^{*}_2(h)= \mathbf{A}_2(h)\textbf{M}_1=\textbf{H}_{2}(\frac{h}{2})\textbf{H}_{1}(h)\textbf{H}_{2}(\frac{h}{2})\textbf{M}_1.
\end{eqnarray}
From the current $n^\text{th}$ iteration to the subsequent $(n+1)^\text{th}$ iteration, the numerical approximations of the dynamical states are given by:
\begin{eqnarray}
\left(\begin{array}{cccc}
\textbf{r}\\
\widetilde{\textbf{r}}\\
\textbf{p}\\
\widetilde{\textbf{p}}
\end{array}\right)_{n+1}
=\mathbf{C}^{*}_2
\left(\begin{array}{cccc}
\textbf{r}\\
\widetilde{\textbf{r}}\\
\textbf{p}\\
\widetilde{\textbf{p}}
\end{array}\right)_{n}.
\end{eqnarray}

\cite{Liu_2016} proposed a fourth-order explicit extended phase-space scheme that employs a pair of Yoshida's triple product constructs. In contrast, the operator $\mathbf{C}^{*}_2$ incorporates just a single instance of such a product with one correction map, which multiplies the speed of computation. Consequently, the 4th-order extended phase-space algorithm with a correction map can be outlined as
\begin{eqnarray}
\textbf{C}_{4}(h)=\textbf{M}_1\otimes\textbf{A}_3(h).\label{eq:6}
\end{eqnarray}

The scale factor $\alpha$ in the expression for $\textbf{C}_{4}$ is solved by leveraging the mean energy values of both $H_1$ and $H_2$,
\begin{eqnarray}
       H(\frac{\textbf{r}+\widetilde{\textbf{r}}}{2},\alpha(\textbf{p}+\widetilde{\textbf{p}})) =\frac{H_{1}(\textbf{r},\widetilde{\textbf{p}})+H_{2}(\widetilde{\textbf{r}},\textbf{p})}{2}. \label{eq:9}
\end{eqnarray}

 The triple-leapfrog product within $\textbf{A}_3(h)$ is defined as $\textbf{A}_{2}(\lambda_{3} h)\textbf{A}_{2}(\lambda_2 h)\textbf{A}_{2}(\lambda_1 h)$, where $\otimes$ signifies the Kronecker product operation. The time coefficients $\lambda_{1},\lambda_{2}$, and $\lambda_{3}$ strictly adhere to the specifications detailed in \cite{Yoshida_1990PRD}. To guarantee fourth-order precision, the combined third-order error terms associated with $\textbf{A}_{2}$ must vanish, mathematically expressed as $\lambda_{1}^{3}+\lambda_{2}^{3}+\lambda_{3}^{3}=0$. Furthermore, these time coefficients must cumulatively amount to a single time step, hence $\lambda_{1}+\lambda_{2}+\lambda_{3}=1$. With these two equations and three unknowns, we opt for the simplifying assumption that $\lambda_{1}=\lambda_{3}$. This yields the time coefficients $\lambda_{1}=\lambda_{3}=1/(2-2^{1/3})$ and $\lambda_{2}=1-2\lambda_{1}$.

The numerical representation spanning the transition from the $n^\text{th}$ to the $(n+1)^\text{th}$ step is accordingly
\begin{align}
\begin{split}
&\textbf{r}_{n+\frac{1}{6}}=\textbf{r}_{n}+\frac{\lambda_1h}{2}\nabla_{\textbf{p}}H_{2}(\widetilde{\textbf{r}}_{n},\textbf{p}_{n})\nonumber\\
&\widetilde{\textbf{p}}_{n+\frac{1}{6}}=\widetilde{\textbf{p}}_{n}-\frac{\lambda_1h}{2}\nabla_{\widetilde{\textbf{r}}}H_{2}(\widetilde{\textbf{r}}_{n},\textbf{p}_{n})\nonumber\\
&\widetilde{\textbf{r}}_{n+\frac{2}{6}}=\widetilde{\textbf{r}}_{n}+\lambda_1h\nabla_{\widetilde{\textbf{p}}}H_{1}(\textbf{r}_{n+\frac{1}{6}},\widetilde{\textbf{p}}_{n+\frac{1}{6}})\nonumber\\
&\textbf{p}_{n+\frac{2}{6}}=\textbf{p}_{n}-\lambda_1h\nabla_{\textbf{r}}H_{1}(\textbf{r}_{n+\frac{1}{6}},\widetilde{\textbf{p}}_{n+\frac{1}{6}})\nonumber\\
&\widetilde{\textbf{p}}_{n+\frac{2}{6}}=\widetilde{\textbf{p}}_{n+\frac{1}{6}}-\frac{\lambda_1h}{2}\nabla_{\widetilde{\textbf{r}}}H_{2}(\widetilde{\textbf{r}}_{n+\frac{2}{6}},\textbf{p}_{n+\frac{2}{6}})\nonumber\\
&\textbf{r}_{n+\frac{2}{6}}=\textbf{r}_{n+\frac{1}{6}}+\frac{\lambda_1h}{2}\nabla_{\textbf{p}}H_{2}(\widetilde{\textbf{r}}_{n+\frac{2}{6}},\textbf{p}_{n+\frac{2}{6}})\nonumber\\
&\textbf{r}_{n+\frac{3}{6}}=\textbf{r}_{n+\frac{2}{6}}+\frac{\lambda_2h}{2}\nabla_{\textbf{p}}H_{2}(\widetilde{\textbf{r}}_{n+\frac{2}{6}},\textbf{p}_{n+\frac{2}{6}})\nonumber\\
&\widetilde{\textbf{p}}_{n+\frac{3}{6}}=\widetilde{\textbf{p}}_{n+\frac{2}{6}}-\frac{\lambda_2h}{2}\nabla_{\widetilde{\textbf{r}}}H_{2}(\widetilde{\textbf{r}}_{n+\frac{2}{6}},\textbf{p}_{n+\frac{2}{6}})\nonumber\\
&\widetilde{\textbf{r}}_{n+\frac{4}{6}}=\widetilde{\textbf{r}}_{n+\frac{2}{6}}+\lambda_2h\nabla_{\widetilde{\textbf{p}}}H_{1}(\textbf{r}_{n+\frac{3}{6}},\widetilde{\textbf{p}}_{n+\frac{3}{6}})\nonumber\\
&\textbf{p}_{n+\frac{4}{6}}=\textbf{p}_{n+\frac{2}{6}}-\lambda_2h\nabla_{\textbf{r}}H_{1}(\textbf{r}_{n+\frac{3}{6}},\widetilde{\textbf{p}}_{n+\frac{3}{6}})\nonumber\\
&\widetilde{\textbf{p}}_{n+\frac{4}{6}}=\widetilde{\textbf{p}}_{n+\frac{3}{6}}-\frac{\lambda_2h}{2}\nabla_{\widetilde{\textbf{r}}}H_{2}(\widetilde{\textbf{r}}_{n+\frac{4}{6}},\textbf{p}_{n+\frac{4}{6}})\nonumber\\
&\textbf{r}_{n+\frac{4}{6}}=\textbf{r}_{n+\frac{5}{6}}+\frac{\lambda_2h}{2}\nabla_{\textbf{p}}H_{2}(\widetilde{\textbf{r}}_{n+\frac{4}{6}},\textbf{p}_{n+\frac{4}{6}})\nonumber\\
&\textbf{r}_{n+\frac{5}{6}}=\textbf{r}_{n+\frac{4}{6}}+\frac{\lambda_3h}{2}\nabla_{\textbf{p}}H_{2}(\widetilde{\textbf{r}}_{n+\frac{4}{6}},\textbf{p}_{n+\frac{4}{6}})\nonumber\\
&\widetilde{\textbf{p}}_{n+\frac{5}{6}}=\widetilde{\textbf{p}}_{n+\frac{4}{6}}-\frac{\lambda_3h}{2}\nabla_{\widetilde{\textbf{r}}}H_{2}(\widetilde{\textbf{r}}_{n+\frac{4}{6}},\textbf{p}_{n+\frac{4}{6}})\nonumber\\
&\widetilde{\textbf{r}}_{n+1}=\widetilde{\textbf{r}}_{n+\frac{4}{6}}+\lambda_3h\nabla_{\widetilde{\textbf{p}}}H_{1}(\textbf{r}_{n+\frac{5}{6}},\widetilde{\textbf{p}}_{n+\frac{5}{6}})\nonumber\\
&\textbf{p}_{n+1}=\textbf{p}_{n+\frac{4}{6}}-\lambda_3h\nabla_{\textbf{r}}H_{1}(\textbf{r}_{n+\frac{5}{6}},\widetilde{\textbf{p}}_{n+\frac{5}{6}})\nonumber\\
&\widetilde{\textbf{p}}_{n+1}=\widetilde{\textbf{p}}_{n+\frac{5}{6}}-\frac{\lambda_3h}{2}\nabla_{\widetilde{\textbf{r}}}H_{2}(\widetilde{\textbf{r}}_{n+1},\textbf{p}_{n+1})\nonumber\\
&\textbf{r}_{n+1}=\textbf{r}_{n+\frac{5}{6}}+\frac{\lambda_3h}{2}\nabla_{\textbf{p}}H_{2}(\widetilde{\textbf{r}}_{n+1},\textbf{p}_{n+1})\nonumber\\
&\alpha=solve[Eq.\ref{eq:9},\alpha]\nonumber\\
\end{split}
\end{align}

\begin{align}
&\textbf{r}=\frac{\textbf{r}_{n+1}+\widetilde{\textbf{r}}_{n+1}}{2},\nonumber\\
&\textbf{r}_{n+1}=\widetilde{\textbf{r}}_{n+1}=\textbf{r}\nonumber\\
&\textbf{p}=\frac{\alpha(\textbf{p}_{n+1}+\widetilde{\textbf{p}}_{n+1})}{2},\nonumber \\
&\textbf{p}_{n+1}=\widetilde{\textbf{p}}_{n+1}=\textbf{p}\label{eq:10}
\end{align}

The final solutions are given by 

\begin{eqnarray}
\left(\begin{array}{cccc}
\textbf{r}^{*}\\
\widetilde{\textbf{r}}^{*}\\
\textbf{p}^*\\
\widetilde{\textbf{p}}^*
\end{array}\right)
=
\left(\begin{array}{cccc}
\frac{(\mathbf{r}+\widetilde{\textbf{r}})}{2}\\
\frac{(\mathbf{r}+\widetilde{\textbf{r}})}{2}\\
\frac{\alpha (\mathbf{p}+\widetilde{\textbf{p}})}{2}\\
\frac{\alpha (\mathbf{p}+\widetilde{\textbf{p}})}{2}
\end{array}\right).
\end{eqnarray}

The correction map devised by \cite{Luo_2020} exhibits several distinctive attributes that surpass its predecessors. Not only does it ensure the preservation of equivalence between the original dynamical variables and their replicated counterparts, but it also uniquely guarantees that the value of the modified Hamiltonian $\widetilde{H}$ remains unaltered following the application of the correction map – a feat unattainable by prior mapping strategies. As a result, the extended phase-space methodology fortified with this innovative correction map demonstrates exceptional performance characterized by elevated efficiency, robust stability, and superior accuracy.

The numerical solutions derived from the algorithms under scrutiny serve as critical variables in the subsequent calculation of gravitational waveforms. It is essential to bear in mind that the actual waveform detected by an observer will be influenced not only by the intrinsic properties of the gravitational waves themselves but also by the relative position and orientation of both the wave source and the observer. Assuming our observer is situated within the $xoz$ plane, the direction of the gravitational wave along the intersection of the orbital plane with the celestial horizon can be denoted by the unit vector $\widehat{\mathbf{p}}=\left(\widehat{p}_x,\widehat{p}_y,\widehat{p}_z\right)$.

To fully capture the observer's perspective, we introduce another unit vector $\widehat{\mathbf{q}}$, which is defined as the cross product of $\widehat{\mathbf{p}}$ and the observer's orientation vector $\widehat{\mathbf{N}}=\left(\widehat{N}_x,\widehat{N}_y,\widehat{N}_z\right)$. These two vectors, $\widehat{\mathbf{p}}$ and $\widehat{\mathbf{q}}$, together form an orthogonal basis that enables us to express the two distinct polarization states of the gravitational wave:

1. Plus-polarization ($h_+$): This component arises from the stretching and squeezing of space perpendicular to the wave's propagation direction. Mathematically, it is given by the following expression:
   \[
   h_{+} = \frac{1}{2}\left(\widehat{p}_i \widehat{p}_j - \widehat{q}_i \widehat{q}_j\right) \widehat{h}^{ij},
   \]
   where the indices $i$ and $j$ run over $x$, $y$, and $z$, and Einstein's summation convention implies a sum over repeated indices. The tensor $\widehat{h}^{ij}$ represents the perturbation in the spacetime metric due to the gravitational wave.

2. Cross-polarization ($h_\times$): This polarization state corresponds to the shear pattern that rotates the orthogonal planes perpendicular to the wave's propagation direction. It is formulated as:
   \[
   h_{\times} = \frac{1}{2}\left(\widehat{p}_i \widehat{q}_j + \widehat{p}_j \widehat{q}_i\right) \widehat{h}^{ij}.
   \]

In essence, $h_+$ and $h_\times$ describe the strain experienced by a freely falling test mass at the observer's location due to the passing gravitational wave, with each component encoding a different aspect of the wave's spatial deformation.

Now, focusing on the gravitational waveform itself, we consider its representation within the framework of the post-Newtonian (PN) approximation. Specifically, the PN approximation of the waveform $\boldsymbol{h}^{i j}$, as developed by Will et al. \cite{will1996gravitational}, are written as:
\begin{eqnarray}
\boldsymbol{h}^{i j}= & \frac{2 \eta M}{D}[\widetilde{\boldsymbol{Q}}^{i j}+\boldsymbol{P}^{0.5} \boldsymbol{Q}^{i j}+\boldsymbol{P} \boldsymbol{Q}^{i j}+\boldsymbol{P}^{1.5} \boldsymbol{Q}^{i j}]_{\mathrm{TT}}
\end{eqnarray}
In this context, $D$ denotes the distance separating the observer from the gravitational wave source. The superscript notation, e.g., $\boldsymbol{P}^{1.5}$, signifies the effective post-Newtonian (PN) order of each term, providing a measure of the relative importance of various relativistic corrections. The subscripts attached to these terms identify their specific physical origin or characteristic.

The individual components of the tensor $\boldsymbol{h}^{i j}$, which describe the spacetime perturbations caused by the gravitational wave, are furnished as follows:
\begin{equation*}
\begin{split}
\widetilde{\boldsymbol{Q}}^{i j}= & 2\left(\boldsymbol{v}^{i} \boldsymbol{v}^{j}-\frac{M}{r} \boldsymbol{n}^{i} \boldsymbol{n}^{j}\right), 
\end{split}
\end{equation*}

\begin{equation}
\begin{split}
\boldsymbol{P}^{0.5} \boldsymbol{Q}^{i j}= & \frac{\delta m}{M}\{3(\widehat{\mathbf{N}} \cdot \mathbf{n}) \frac{M}{r}\left[2 n^{(i} v^{j)}-\dot{R} n^{i} n^{j}\right] \\
& +(\widehat{\mathbf{N}} \cdot \mathbf{v})\left[\frac{M}{r} n^{i} n^{j}-2 v^{i} v^{j}\right]\}
\end{split}
\end{equation}

\begin{equation}
\begin{split}
& \boldsymbol{P} \boldsymbol{Q}^{i j}=\frac{1}{3}\{( 1 - 3 \eta ) [(\widehat{\mathbf{N}} \cdot \mathbf{n})^{2} \frac{M}{r}[(3 v^{2}-15 \dot{R}^{2}+7 \frac{M}{r}) n^{i} n^{j}+\\
& 30 \dot{R} n^{(i} v^{j)}-14 v^{i} v^{j}]+(\widehat{\mathbf{N}} \cdot \mathbf{n})(\widehat{\mathbf{N}} \cdot \mathbf{v})\frac{M}{r}[12 \dot{R} n^{i} n^{j}  \\
&-32 n^{(i} v^{j)}]+(\widehat{\mathbf{N}} \cdot \mathbf{v})^{2}[6 v^{i}v^{j}-2 \frac{M}{r} n^{i} n^{j}]]+[3(1-3 \eta) v^{2}
 \\
& -2(2-3 \eta) \frac{M}{r}] v^{i} v^{j} +4 \frac{M}{r} \dot{R}(5+3 \eta) n^{(i} v^{j)}+\\
& \frac{M}{r}[3(1-3 \eta) \dot{R}^{2}-(10+3 \eta) v^{2}+29 \frac{M}{r}] n^{i} n^{j}\}, 
\end{split}
\end{equation}

\begin{equation}
\begin{split}
&\boldsymbol{P}^{1.5} \boldsymbol{Q}^{i j}=\frac{\delta m}{M}(1-2 \eta)\{(\widehat{\mathbf{N}} \cdot \mathbf{n})^{3} \frac{M}{r}[\frac{5}{4}(3 v^{2}-7 \dot{R}^{2}+\\
& 6 \frac{M}{r}) \dot{R}^{2} n^{i} n^{j}-\frac{17}{2} \dot{R} v^{i} v^{j}-\frac{1}{6}(21 v^{2}-105 \dot{R}^{2}+44 \frac{M}{r}) n^{(i} v^{j)}] \\
& +\frac{1}{4}(\widehat{\mathbf{N}} \cdot \mathbf{n})^{2}(\widehat{\mathbf{N}} \cdot \mathbf{v}) \frac{M}{r}[58 v^{i} v^{j}+(45 \dot{R}^{2}-9 v^{2}\\
& -28 \frac{M}{r}) n^{i} n^{j}-108 \dot{\boldsymbol{r}} n^{(i} v^{j)}]+\frac{3}{2}(\widehat{\mathbf{N}} \cdot \mathbf{n})(\widehat{\mathbf{N}} \cdot \mathbf{v})^{2} \\
& \frac{M}{r}[10 n^{(i} v^{j)}-3 \dot{R} n^{i} n^{j}]+\frac{1}{2}(\widehat{\mathbf{N}} \cdot \mathbf{v})^{3}(\frac{M}{r} n^{i} n^{j}\\
& -4 v^{i} v^{j})\}+\frac{1}{12} \frac{\delta m}{M}(\widehat{\mathbf{N}} \cdot \mathbf{n}) \frac{M}{r}\{2 \boldsymbol { n } ^ { ( i } \boldsymbol { v } ^ { j ) } [\dot{R}^{2}(63 \\
& +54 \eta)-\frac{M}{r}(128-36 \eta)+v^{2}(33-18 \eta)]\\
&+n^{i} n^{j} \dot{R}(\dot{R}^{2}(15-90 \eta)-v^{2}(63-54 \eta) \\
& +\frac{M}{r}(242-24 \eta))-\dot{R} v^{i} v^{j}(186+24 \eta)\}+\\
& \frac{\delta m}{M}(\widehat{\mathbf{N}} \cdot \mathbf{v})\{\frac{1}{2} v^{i} v^{j}[\frac{M}{r}(3-8 \eta)-2 v^{2}(1-5 \eta)] \\
& -n^{(i} v^{j)} \frac{M}{r} \dot{R}(7+4 \eta)-n^{i} n^{j} \frac{M}{r}[\frac{3}{4}(1-2 \eta) \dot{R}^{2}\\
& +\frac{1}{3}(26-3 \eta) \frac{M}{r}-\frac{1}{4}(7-2 \eta)]\}
\end{split}
\end{equation}
In the expressions above, $\delta m=m_1-m_2$, $\mathbf{v}=(v^1, v^2, v^3)=(1+\frac{1}{2}(3\eta -1)p^2-\frac{(3+\eta)}{r})\mathbf{p}-\eta(\mathbf{n} \cdot \mathbf{p})\frac{n}{r}$, where $\dot{R} = \mathbf{n} \cdot \mathbf{v}$ represents the radial velocity component.

Upon obtaining numerical solutions for the time-evolution of the position vector $\mathbf{r}$ and the momentum vector $\mathbf{p}$ of a spinning compact binary system through the application of suitable numerical methods, we can proceed to construct the corresponding gravitational waveforms. These waveforms encode the distortions in spacetime induced by the binary's orbital dynamics, providing a theoretical template against which real gravitational wave detections can be compared and analyzed. 

Since the key problem is to predict the time-evolution of the position vector $\mathbf{r}$ and the momentum vector $\mathbf{p}$ of a spinning compact binary system, the machine learning methods can help. What is needed is a data set generated from the numerical solutions mentioned above. The corresponding data generations of algorithm $\textbf{C}_4(h)$ are as the following.

\begin{algorithm}
\raggedright
\caption{Given initial celestial coordinates and momenta \([r, p]_0\), apply symplectic-like integration to compute the numerical evolution of the celestial coordinates \([r, p]_{n*\Delta t}\) at time intervals \(\Delta t\) within the time interval \([0,t]\) using the symplectic-like integrator \(\textbf{C}_{4}(h)\). Here, \(n=1, 2,...,\frac{t}{\Delta t}\) and \(t\) is a multiple of \(\Delta t\).}
\label{algo:symplectic}
\KwIn{Initial conditions: \([r, p]_n\)}
\BlankLine
Copy variables \([r, p]\) to \([r, \widetilde{r}, p, \widetilde{p}]_n\) to extend phase space, where \([r, \widetilde{r}, p, \widetilde{p}]_n\) are the new input variables for the symplectic integrator \(\textbf{C}_{4}(h)\); \\
Initialize integration step \(n=0\); \\
Initialize time step \(\Delta t=h\); \\
Initialize simulation time \(t=0\); \\
\While{\(t < t_{max}\)}{
  Perform symplectic-like integration using \(\textbf{C}_{4}(h)\); \\
  Compute the new position \(r\) and momentum \(p\) for the next integration step using \(\textbf{C}_{4}(h)\); \\
  
  \begin{equation}
  \left(\begin{array}{c}
  r_{n+1} \\
  \widetilde{r}_{n+1} \\
  p_{n+1} \\
  \widetilde{p}_{n+1}
  \end{array}\right) = \textbf{C}_{4}(h)
  \left(\begin{array}{c}
  r_{n} \\
  \widetilde{r}_{n} \\
  p_{n} \\
  \widetilde{p}_{n}
  \end{array}\right);
  \end{equation}
  Output: \([r, \widetilde{r}, p,\widetilde{p}]_{n+1}\); \\
  Increment integration step \(n = n + 1\); \\
  Update current time to \(t = n * \Delta t\); \\
}
\KwOut{Final celestial coordinates and momenta \([r, p]_{n+1}\)}
\end{algorithm}

Hence, utilizing the Algorithm \ref{algo:symplectic}, we generated a comprehensive dataset comprising numerical solutions that describe the evolutionary orbits of compact binaries.
These high-fidelity simulations serve as an invaluable resource, constituting a dataset suitable for application in deep learning models.
The dataset for the two models are divided in the following way: 80\% for training, 10\% for validation and 10\% for testing.
In subsequent sections, we will indeed employ these data to train and potentially enhance our predictive capabilities in understanding and modeling the intricate dynamics of such astrophysical systems.
\section{\label{sec3} Multivariate time series with deep networks}

\subsection{\label{sec31} Deep network architectures}

Unlike univariate time series forecasting, multivariate time series forecasting is more challenging \citep{beeram2020time}.
However, one benefit of multivariate time series forecasting is its ability to not only effectively predict complex systems with interrelated variables, but to also offer additional information with variables for deep network models.
The intricate and multidimensional nature of multivariate time series data poses challenges for traditional time series forecasting models like autoregressive integrated moving average (ARIMA) \citep{box1970distribution}.
ARIMA's limitations become especially apparent in handling such data as it assumes linear interactions between variables, an oversimplified assumption for complex astronomical datasets.
Therefore, we investigate the utilization of advanced and complex deep learning models for predicting multivariate time series and apply two kinds of models in this time series forecasting task.

LSTM is one of the most well-used sequential model in various time series tasks, and TCN is a convolutional model that can learn the representation of 1-Dimensional sequential data.
By applying these two deep learning models, we aim to provide a comprehensive understanding of the properties and challenge exist in this work.

\begin{table*}
\caption{Performance metrics of LSTM models with different architecture configurations.}
\label{tab:lstm_performance}
\centering
\setlength{\tabcolsep}{6mm}{
\begin{tabular}{ccccccc}
\hline
\hline
\textbf{Model}        & \textbf{Layers} & \textbf{Hidden Units} & \textbf{Window Size} & \textbf{RMSE $\downarrow$} & \textbf{$R^2$ $\uparrow$} \\
\hline
\multirow{6}{*}{LSTM} & 2               & 256                   & 200                  & 0.0175        & 0.9960     \\
                      & 2               & 512                   & 200                  & 0.0336        & 0.9859      \\ \cline{2-6} 
                      & 3               & 256                   & 200                  & 0.0160        & 0.9965      \\
                      & 3               & 512                   & 200                  & \textbf{0.0138}        & \textbf{0.9972}      \\ \cline{2-6} 
                      & 4               & 256                   & 200                  & 0.0244        & 0.9918      \\
                      & 4               & 512                   & 200                  & 0.0235        & 0.9928      \\
\hline
\hline
\end{tabular}
}
\end{table*}

\begin{table*}
\caption{Performance metrics of TCN models with different architecture configurations.}
\label{tab:tcn_performance}
\centering
\setlength{\tabcolsep}{6mm}{
\begin{tabular}{ccccccc}
\hline
\hline
\textbf{Model}        & \textbf{Kernel Size} & \textbf{Filters} & \textbf{Window Size} & \textbf{RMSE $\downarrow$} & \textbf{$R^2$ $\uparrow$} \\
\hline
\multirow{9}{*}{TCN} & 3               & 128                   & 200    & 0.0452        & 0.9746      \\                           & 3               & 256                   & 200    & 0.0447        & 0.9760      \\
                      & 3               & 512                   & 200   & 0.0414        & 0.9795      \\ \cline{2-6} 
                      & 5               & 128                   & 200   & 0.0318        & 0.9875      \\
                      & 5               & 256                   & 200   & \textbf{0.0301}        & \textbf{0.9879}      \\
                      & 5               & 512                   & 200   & 0.0367        & 0.9831      \\ \cline{2-6} 
                      & 7               & 128                   & 200   & 0.0423        & 0.9766      \\
                      & 7               & 256                   & 200   & 0.0351        & 0.9860      \\
                      & 7               & 512                   & 200   & 0.0399        & 0.9803      \\
\hline
\hline
\end{tabular}
}
\end{table*}

\subsubsection{\label{sec311} Long Short-Term Memory}

\begin{figure}
    \centering
    \includegraphics[width=\linewidth]{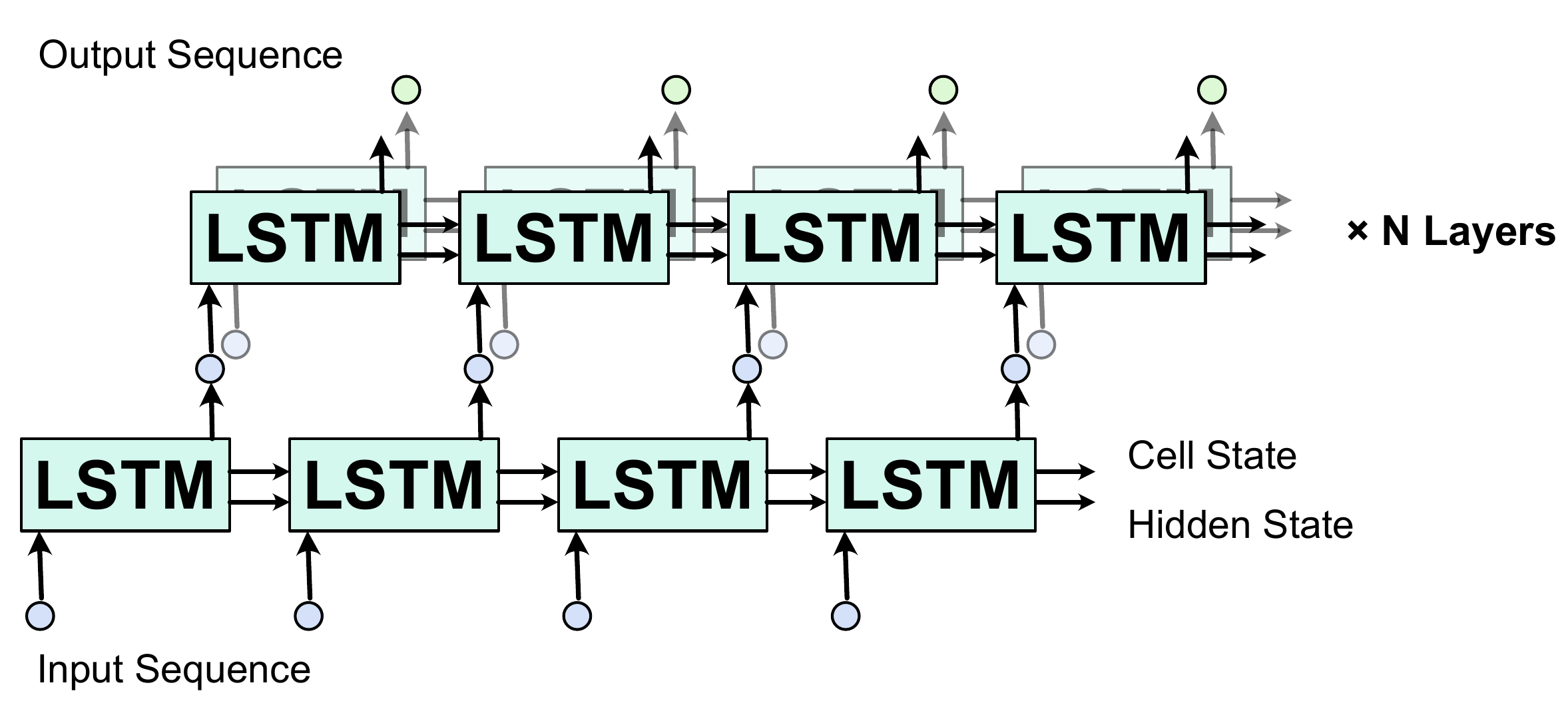}
    \caption{The network architecture of multi-layer LSTM in our experiments.}
    \label{fig:lstm-arch}
\end{figure}

LSTM, an improved version of recurrent neural network (RNN), learns the long-term dependencies by having unique memory state design as shown in Figure \ref{fig:lstm-arch}.
It processes data by looping through the entire sequence.
In our experiment, LSTM processes features at a single moment sequentially from the start time to the end time.
It persists the information it previously seen in its memory at each moment.
The memory is in the form of cell state and hidden state, which are the vector outputs from the model.
These memory information together with the data from the next moment are used as inputs of next loop.
This design helps LSTM learn the pattern of time evolution from previous moments, and converts the pattern information into feature representation in both hidden state and cell state.

Our LSTM network is stacked by multiple LSTM layers as shown in Figure \ref{fig:lstm-arch}, where the output of one layer becomes the input to the next.
This structure allows the network to learn more complex features at multiple levels of abstraction.

\subsubsection{\label{sec312} Temporal Convolutional Network}

\begin{figure*}
    \centering
    \includegraphics[width=\linewidth]{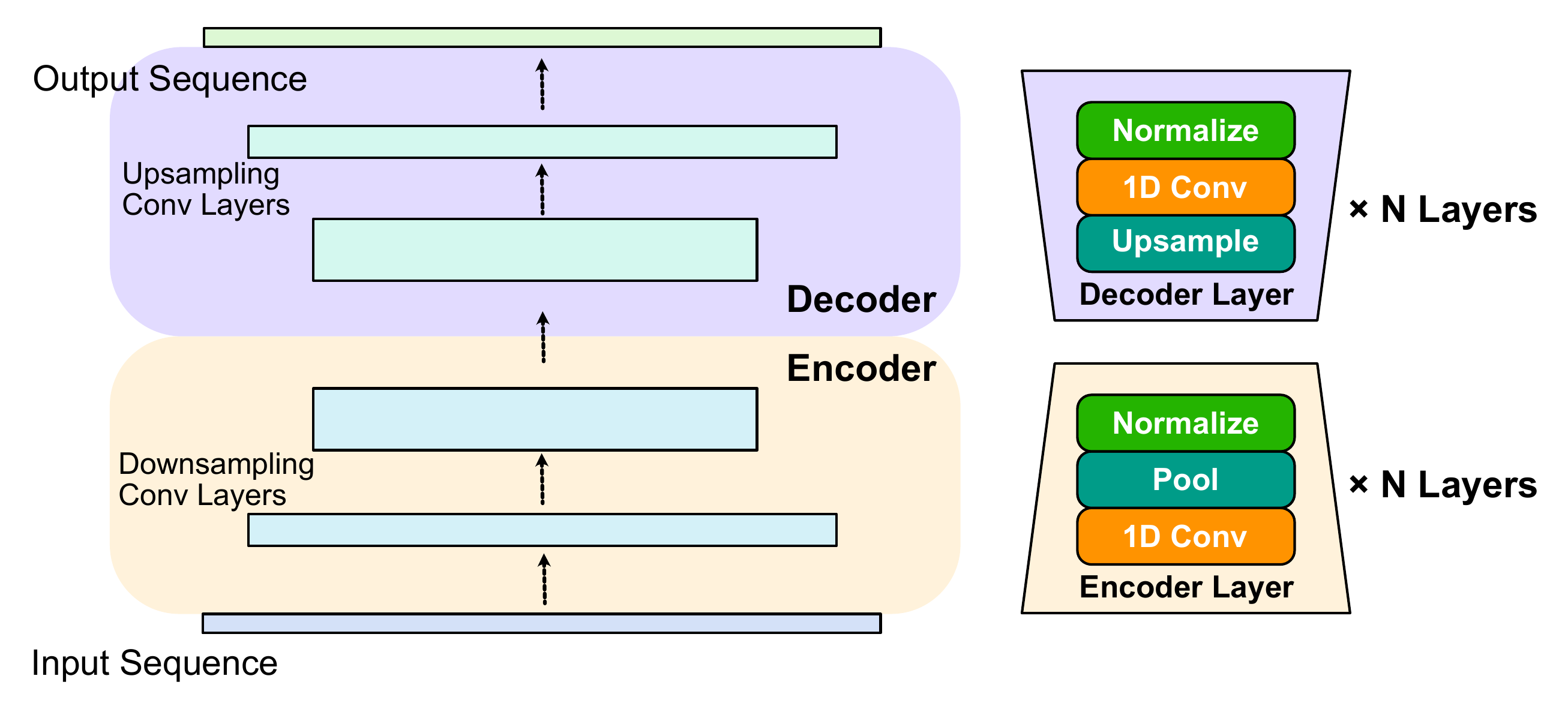}
    \caption{Illustration of the TCN network architecture in our experiments.}
    \label{fig:tcn-arch}
\end{figure*}

TCN \citep{lea2017temporal} is a convolutional neural network (CNN) that focuses on time series problems.
The design of TCN differs significantly from that of LSTM since TCN captures temporal dependencies by using convolutional layers instead of memory units.
Meanwhile, TCN implements causal convolutional layers, which offer greater flexibility in handling inputs of varying lengths.

The architecture of TCN is shown in Figure \ref{fig:tcn-arch}.
It has the advantages of a typical CNN and has achieved great success in image-related tasks.
TCN is the equivalent of CNN, which takes 1-dimensional sequential data as input.
For each layer of TCN, there are many 1-dimensional convolutions that learn the low level feature over time.
Then, the pooling operation is applied to capture the temporal patterns.
The stack of convolutions and the pooling operations enables TCN to hierarchically capture different range of pattern and dependency.
Given this ability, TCNs learn long-term dependencies better than other time series forecasting networks \citep{hewage2020temporal, gopali2021comparison}.
In LSTM, the previous information are only stored in its cell state and hidden state.
TCN takes previous raw data across a much longer period of time, which gives it richer information related to the context and the data pattern in forecasting the upcoming sequence.

\subsubsection{\label{sec313} Tuning hyper-parameters of network architectures}
In this study, we identify the optimal network architecture for both LSTM and TCN models, tailored to our specific dataset.
We experiment with varying numbers of layers and dimensions of hidden units for the LSTM model.
In contrast, for the TCN model, we explore different kernel sizes and quantities of filters.
Both models employ a window size of 200 steps for time series prediction.
The performance metrics of the LSTM and TCN models, under diverse architectural configurations, are described in Tables \ref{tab:lstm_performance} and \ref{tab:tcn_performance}, respectively.

As shown in Table \ref{tab:lstm_performance}, we consider LSTM networks with 2, 3, and 4 layers, and configurations featuring 256 and 512 hidden units.
Table \ref{tab:lstm_performance} reveals that the LSTM with 2 layers is underfitting, particularly evident in configurations with 512 hidden units.
Furthermore, our findings consistently demonstrate that no matter how many hidden units are used, the results of 3-layer LSTM always outperform those of 2- and 4-layer architectures.
After comparison, the optimal performance is obtained when using 3 layers of LSTM and 512 hidden units.

In order to find the optimal hyper-parameters of TCN, we consider kernel size of 3,5,7 and the number of 128, 256, and 512 filters for performance comparison.
The larger kernel size on one hand provides a larger sensory field and stronger feature extraction capability, but on the other hand, it also appears to ignore some subtle local features.
Similarly, more filters can improve the representation ability of the model, but may also lead to overfitting.
The purpose of this experiment is to find the most suitable architecture configurations for our dataset.
The results in Table \ref{tab:tcn_performance} show that the performance of TCN is optimized when the kernel size is equal to 5 and 256 filters are used.
When the kernel size is equal to 5 and 7, too large filters will cause overfitting, which leads to performance degradation.

\subsection{\label{sec32} Deep Neural Network Optimization with Physical Conditions}

Several physical conditions, including the Hamiltonian conservation, the angular momentum conservation, and the canonical equations, have been considered in order to improve the machine learning models.
They have been included in the loss functions and tested to see if they could improve the machine learning models.
It is important to note that our approach differs from PINNs in that PINNs employ physical equations as solvers, whereas we utilize physical equations as constraints.
Additionally, the previous study \citep{hu2020neural} introduced a method named Neural-PDE, which uses LSTM networks to learn and predict the numerical solutions of Partial Differential Equations (PDEs).
They found that LSTM based neural network can efficiently find the solutions to multidimensional PDEs without knowing the specific form of PDE.
This is closely aligned with our approach, where our focus is on utilizing deep learning for sequence predictions, whereas their work concentrated on numerical solutions using deep learning.
In this section, our experimental results demonstrate that employing both LSTM and TCN as black-box predictors for sequence prediction can also significantly improve the prediction performance without using PINN as PDE solvers.

\subsubsection{\label{sec321} Physical conditions in optimization}

\paragraph{Hamiltonian conservation}

The 2PN Hamiltonian $H$ in Equation \ref{2pnhami} is conserved throughout the time series.
The computed $H(\hat{y}_t)$ for each moment is expected equivalent to the initiation $H({x_0})$ at the start of the sequence.
This can be formed as a constraint to the optimization of the loss function. 
\begin{equation}
\label{hamiltonian}
\sum_{t=0}^{T} (H(x_0) - H(\hat{y}_t))^2 = 0 \end{equation}

\paragraph{Angular momentum conservation}

Similar to the Hamiltonian, the angular momentum also follows the conservation law.
The angular momentum can be expressed as the following:

\begin{equation}
\label{angular_momentum}
 A = \mathbf{r} \times \mathbf{p}
\end{equation}

Hence, we have the following constraint.

\begin{equation}
    \sum_{t=0}^{T} (A(x_0) - A(\hat{y}_t))^2 = 0,
\end{equation}

where $A(x)$ is the function in Equation \ref{angular_momentum} to compute angular momentum in given time.
It retrieves $\mathbf{r}, \mathbf{p}$ from the input $x$.

\paragraph{Canonical equations}
The canonical equations are differential equations that can be numerically solved and gets the approximate prediction of the system.

\begin{equation}
\label{he_equ}
    \begin{aligned}
        \frac{d \mathbf{q}}{d t} = \frac{\partial H}{\partial \mathbf{p}}\\
        \frac{d \mathbf{p}}{d t} = - \frac{\partial H}{\partial \mathbf{q}}
    \end{aligned}
\end{equation}

The gradient values of left and right sides of Equation \ref{he_equ} are computed.
Then the following constraint is obtained:

\begin{equation}
    \begin{aligned}
        \min_{W} \quad & \sum_{t=0}^{T} L(y_t, \hat{y}_t; W) \\
        s.t. \quad & \sum_{t=0}^{T} (H(x_0) - H(\hat{y}_t))^2 = 0 \\
        & \sum_{t=0}^{T} (A(x_0) - A(\hat{y}_t))^2 = 0 \\
        & \frac{d \mathbf{q}}{d t} - \frac{\partial H}{\partial \mathbf{p}} = 0 \\
        & \frac{d \mathbf{p}}{d t} + \frac{\partial H}{\partial \mathbf{q}} = 0
    \end{aligned}
\end{equation}

$W$ is the parameter matrix of the model to be optimized and $L$ is the loss function of the model.
The equality constrains are eliminated by optimizing the Lagrangian of the above problem:

\begin{equation}
    \begin{aligned}
        \min_{W} \quad & \mu_1 \sum_{t=0}^{T} L(y_t, \hat{y}_t; W) \\
       & + \mu_2 \sum_{t=0}^{T} (H(x_0) - H(\hat{y}_t))^2 \\
       & + \mu_3 \sum_{t=0}^{T} (A(x_0) - A(\hat{y}_t))^2 \\
       & + \mu_4 \| \frac{d \mathbf{q}}{d t} - \frac{\partial H}{\partial \mathbf{p}}\|_2^2 \\
       & + \mu_5 \| \frac{d \mathbf{p}}{d t} + \frac{\partial H}{\partial \mathbf{q}}\|_2^2,
    \end{aligned}
\end{equation}

In the equation above, the parameters $\mu_1$, $\mu_2$, $\mu_3$, $\mu_4$, and $\mu_5$ serve as hyperparameters, each calibrating the relative importance of the corresponding term in the loss function.
These hyperparameters are tuned to balance the contribution of task-specific loss, Hamiltonian conservation, angular momentum conservation, and the adherence to canonical equations, ensuring an optimized trade-off between model accuracy and physical fidelity.
The sum of hyperparameters $\mu_1$, $\mu_2$, $\mu_3$, $\mu_4$, and $\mu_5$ is constrained to equal 1, ensuring a normalized weighting scheme across all terms.
Predominantly, $\mu_1$ holds the greatest weight, reflecting its primary role in guiding the model's gradient descent for task-specific performance. In contrast, $\mu_2$, $\mu_3$, $\mu_4$, and $\mu_5$ primarily function to impose physical constraints, rather than directly steering the convergence process. This design choice is predicated on the premise that while adherence to physical laws is critical, the primary objective of the model remains task-oriented optimization, necessitating a higher weighting for $\mu_1$.

The final loss function in our study can be constructed as follows:

\begin{equation}\begin{aligned}
\label{equ_loss}
    L(\hat{\mathbf{y}}, \mathbf{y} ; \mathbf{x} ) = & \mu_1 \sum_{t=0}^{T} L(y_t, \hat{y}_t; W) + \\
    & \mu_2 \sum_{t=0}^{T} (H(x_0) - H(\hat{y}_t))^2 + \\
    &\mu_3 \sum_{t=0}^{T} (A(x_0) - A(\hat{y}_t))^2 + \\
    &\mu_4 \| \frac{d \mathbf{q}}{d t} - \frac{\partial H}{\partial \mathbf{p}}\|_2^2 + \\
    &\mu_5 \| \frac{d \mathbf{p}}{d t} + \frac{\partial H}{\partial \mathbf{q}}\|_2^2
\end{aligned}\end{equation}

\begin{table*}[htp]
\caption{Comparative analysis of ablation studies on TCN and LSTM models with progressive inclusion of constraints. The Log-Cosh is defined in Eq.~\ref{logcosh}. The Hamiltonian formula can be found in Eq.~\ref{hamiltonian}. 
The Angular momentum formula is from Eq.~\ref{angular_momentum}.
The canonical equations are presented in Eq.~\ref{he_equ}. 
The hyperparameters ($\mu_1$, $\mu_2$, $\mu_3$, $\mu_4$, and $\mu_5$) are presented in Eq.~\ref{equ_loss}. 
The optimal results are highlighted in red.
}
\label{tab:loss_performance}
\centering
\setlength{\tabcolsep}{3.3mm}{
\begin{tabular}{ccccccc}
\hline
\hline
\textbf{Model}        & \textbf{Experiment} & \textbf{Loss Function} & \textbf{Parameters} & \textbf{RMSE $\downarrow$} & \textbf{$R^2$ $\uparrow$} \\
\hline
\hline
\multirow{20}{*}{TCN} & A  & Log-Cosh & $\mu_1=1$ & 0.0288  & 0.9894 \\
\cline{2-6}
& A+B  & \makecell[c]{Log-Cosh, \\ Hamiltonian} & $\mu_1=1-\mu_2$  & 0.0330 & 0.9865 \\
\cline{2-6}
& A+C  & \makecell[c]{Log-Cosh, \\ Angular momentum} & $\mu_1=1-\mu_3$  & 0.0290  & 0.9890  \\
\cline{2-6}
& A+D  & \makecell[c]{Log-Cosh, \\ canonical equations} & $\mu_1=1-\mu_4-\mu_5$  & \textbf{0.0279} & \textbf{0.9899} \\
\cline{2-6}
& A+B+C  & \makecell[c]{Log-Cosh, \\ Hamiltonian, \\ Angular momentum} & $\mu_1=1-\mu_2-\mu_3$  & 0.0356 & 0.9833 \\
\cline{2-6}
& A+B+D  & \makecell[c]{Log-Cosh, \\ Hamiltonian, \\ canonical equations} & $\mu_1=1-\mu_2-\mu_4-\mu_5$  & \textbf{0.0281} & \textbf{0.9896} \\
\cline{2-6}
& A+C+D  & \makecell[c]{Log-Cosh, \\ Angular momentum, \\ canonical equations} & $\mu_1=1-\mu_3-\mu_4-\mu_5$  & 0.0304 & 0.9882 \\
\cline{2-6}
& A+B+C+D  & \makecell[c]{Log-Cosh, \\ Hamiltonian, \\ Angular momentum, \\ canonical equations} & $\mu_1=1-\mu_2-\mu_3-\mu_4-\mu_5$  & \textbf{\color{red}{0.0256}} & \textbf{\color{red}{0.9919}} \\
\hline
\hline
\multirow{20}{*}{LSTM} & A  & Log-Cosh & $\mu_1=1$ & 0.0138 & 0.9972 \\
\cline{2-6}
& A+B  & \makecell[c]{Log-Cosh, \\ Hamiltonian} & $\mu_1=1-\mu_2$  & 0.0149 & 0.9967 \\
\cline{2-6}
& A+C  & \makecell[c]{Log-Cosh, \\ Angular momentum} & $\mu_1=1-\mu_3$  & 0.0158 & 0.9964  \\
\cline{2-6}
& A+D  & \makecell[c]{Log-Cosh, \\ canonical equations} & $\mu_1=1-\mu_4-\mu_5$  & 0.0150 & 0.9969 \\
\cline{2-6}
& A+B+C  & \makecell[c]{Log-Cosh, \\ Hamiltonian, \\ Angular momentum} & $\mu_1=1-\mu_2-\mu_3$  & 0.0210 & 0.9944 \\
\cline{2-6}
& A+B+D  & \makecell[c]{Log-Cosh, \\ Hamiltonian, \\ canonical equations} & $\mu_1=1-\mu_2-\mu_4-\mu_5$  & 0.0157 & 0.9966 \\
\cline{2-6}
& A+C+D  & \makecell[c]{Log-Cosh, \\ Angular momentum, \\ canonical equations} & $\mu_1=1-\mu_3-\mu_4-\mu_5$  & 0.0148 & 0.9970 \\
\cline{2-6}
& A+B+C+D  & \makecell[c]{Log-Cosh, \\ Hamiltonian, \\ Angular momentum, \\ canonical equations} & $\mu_1=1-\mu_2-\mu_3-\mu_4-\mu_5$  & \textbf{\color{red}{0.0135}} & \textbf{\color{red}{0.9974}}\\
\hline
\hline
\end{tabular}
}
\end{table*}

\subsubsection{\label{sec322} Optimizing loss function design through ablation studies}
As shown in Equation \ref{equ_loss}, our loss function integrates four components: Log-Cosh loss for the task, Hamiltonian MSE for energy conservation, angular momentum MSE for rotational invariance, and the adherence to canonical equations.
In this study, we employe distinct loss functions within LSTM and TCN models, utilizing the optimal network architecture configurations identified in Section \ref{sec31}, specifically a 3-layer LSTM with 512 hidden units and a TCN with 5 kernel size and 256 filters.
We also use the second-order derivative feature enhancement method introduced in Section \ref{sec33} for all experiments.
The input window size is set to 200 for LSTM and 300 for TCN.

Log-Cosh \citep{saleh2022statistical} is a famous loss function used in machine learning for regression.
One of the primary advantages of Log-Cosh over MSE is its robustness to outliers.
Log-Cosh loss function can be expressed as Equation \ref{logcosh}:

\begin{eqnarray}
\label{logcosh}
\text{Log-Cosh} & = & \sum_{i=1}^{n} \log(\cosh(\hat{y}_{i} - y_{i})),
\end{eqnarray}

where $\log$ refers to the natural logarithm, $\cosh$ is the hyperbolic cosine function, $n$ is the number of observations, $y_i$ represents the actual values, and $\hat{y}_i$ represents the predicted values. 

To ensure dimensional consistency between the other physical constraint terms and the Log-Cosh loss function, we employed $\mu$ coefficients to scale these terms, aligning their dimensions closely with that of the Log-Cosh.
Except for $\mu_1$, the values of other $\mu$ coefficients are set within the range of $10^{-3}$ to $10^{-5}$.
This strategy is designed to prevent these constraints from excessively affecting the convergence of the Log-Cosh loss function.

Table \ref{tab:loss_performance} presents the results of ablation studies conducted on TCN and LSTM models with the progressive inclusion of constraints.
In the second column of Table \ref{tab:loss_performance}, A, B, C, and D represent the four components of the complete loss function. 
Results that showed improvements over the experiments using only Log-Cosh as the loss function were bolded, with the optimal outcomes highlighted in red.
The experimental results demonstrate that the combined application of the three physical constraints positively influences the training effectiveness of the model.
The incorporation of all physical constraints led to performance enhancements in both TCN and LSTM models. 
However, the improvement extent for LSTM is not as significant as for TCN, possibly due to the LSTM's inherent capability to fit the data distribution well, making the addition of physical constraints less impactful.
For TCN, the results demonstrated that including physical constraints in the loss function effectively compensated for the model's fitting deficiencies, and that even a small constant before the physical constraint terms could aid the training process in converging in the correct direction.

Also, we find that using the Hamiltonian conservation and the angular momentum conservation alone leads to performance degradation.
The ideas of manifold correction methods predominantly focus on conserve certain physical quantities (Hamiltonian conservation or Angular momentum conservation) that should theoretically remain invariant throughout the simulation process, as demonstrated by \cite{Ma:2008ApJ, Wang_2018AAS}.
These strategies generally serve to mitigate the propagation of numerical errors and, consequently, enhance the proximity of the computed numerical solution to the true, analytical solution.
Despite their effectiveness in many instances, the failure of this method is demonstrated in article \cite{Luo_2022}.
Despite the fact that these techniques strive to minimize the deviations in the constants of motion, it is evident that they lead to numerical solutions that diverge from the exact solutions rather than converging towards them.
This observation implies that while these manifold correction methods might be successful in maintaining specific conserved quantities, they may inadvertently introduce additional complexities or inaccuracies that cause the overall numerical trajectories to deviate from the expected theoretical paths in the given test scenarios.
Considering Hamiltonian conservation ensures that the total energy calculation is earned, but the energies of the subterms deviate from their true values.

In contrast, incorporating canonical equations into the loss function contributed to an improvement in model performance.
In the experiments with the LSTM model, even though the result with solely canonical equations (A+D setup) does not surpass the baseline, it is notably close to that achieved with only the Log-Cosh loss function.
Adding canonical equations as a restriction improves the generalization ability of the machine learning training model because essentially this restriction is equivalent to a first-order algorithm, which implies the same physical laws as the numerical solution of the fourth-order algorithm $CM4$, but with a much lower numerical accuracy, and thus if the weighting factors are too high and focus too much on satisfying the regular equations, it will instead reduce the learning accuracy.

\subsection{\label{sec33} Data pre-processing \& Feature augmentation}

\begin{figure*}
\begin{center}
\includegraphics[scale=0.7]{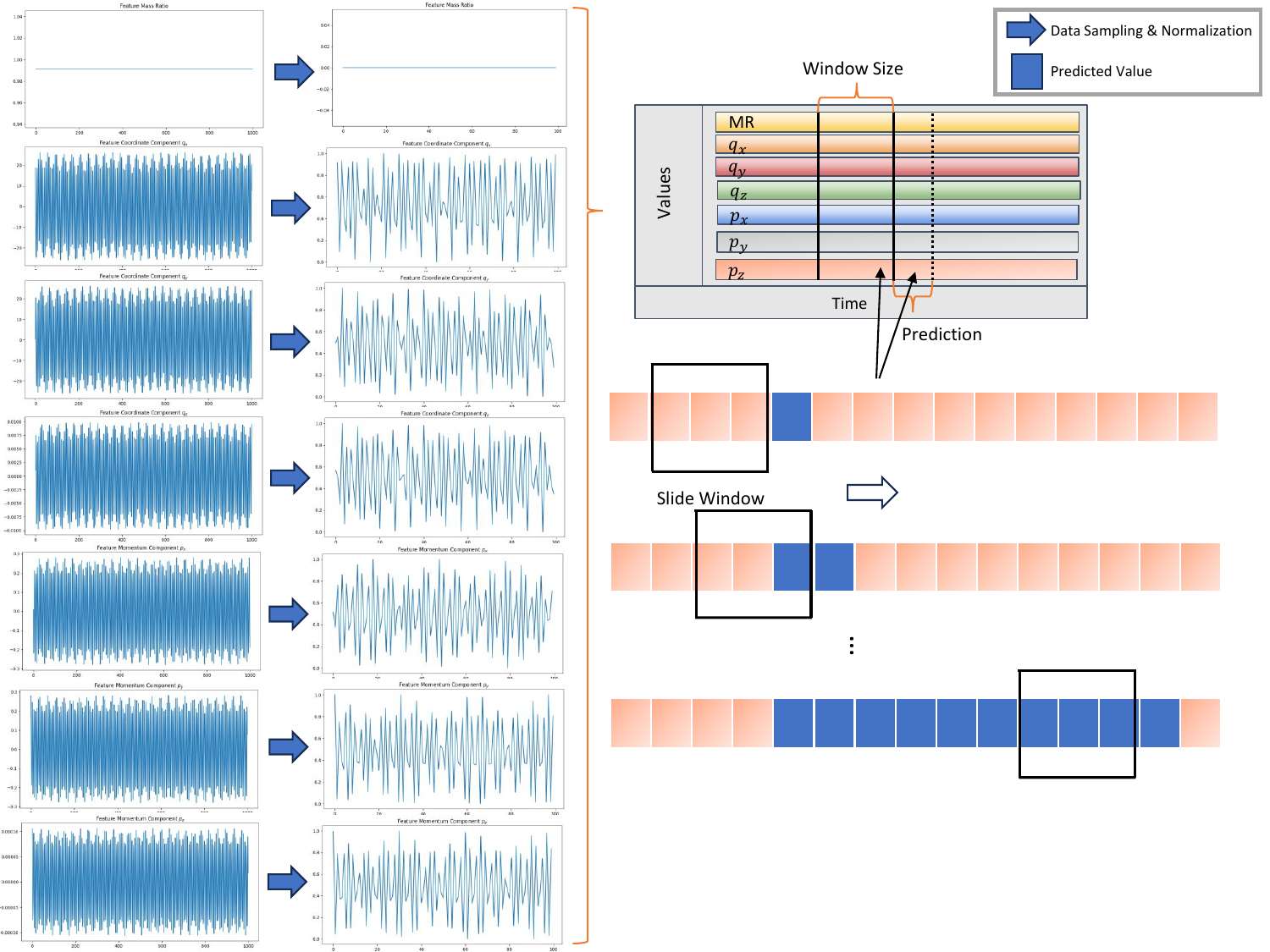}
\caption{Illustration of the data pre-processing and predicting process.}
\label{highlvlidea}
\end{center}
\end{figure*}

This section describes the data pre-processing methods and the data augmentation methods used in the experiments.
Figure \ref{highlvlidea} illustrates the data pre-processing and predicting process.
Due to the issue of gradient vanishing in long sequence prediction observed in RNN-based LSTM models \citep{noh2021analysis}, it becomes imperative to employ uniform sampling on input sequences to reduce the length of sequences.
Considering that too few data points may have an impact on the distribution, we select a reasonable number, uniformly sampling 1000 data points from each sequence.
To eliminate the influence of differences in magnitude and range of values between features, we employ the following min-max normalization formula to normalize the input features:

\begin{eqnarray}
\label{eq_nor}
x_{scaled} & = & \frac{x - x_{min}}{x_{max} - x_{min}},
\end{eqnarray}

where $x_{min}$ and $x_{max}$ are the minimum and maximum value of $x$.
As a result, all the scaled features are in the range of $[0, 1]$ and concatenated to form a multivariate time series of shape 7.

The predicting process also is shown in Figure \ref{highlvlidea}.
During inference for LSTM or TCN models, a stepwise prediction process is employed with a sliding window approach.
We first initial window to predict the value and incorporate the predicted value from the previous step into the input window.
Then, we employ the new input window to make the prediction for the next time step.
These steps are iterated, progressively integrating the most recent predictions into the input window, until the desired number of prediction is achieved.

In order to find the optimal window size, we compare the performance of different window sizes on both LSTM and TCN models.
We utilize a 3-layer LSTM model with 512 hidden units and a TCN network with 256 filters and a kernel size of 5 as the experimental models for our study.
To minimize confounding factors in the experiments, we do not to employ the custom loss function mentioned in Section \ref{sec32}.
We make predictions for 20 time steps and use the same feature augmentation method in both models.
The results of the experiment are shown in Table \ref{windowsize}.
According to the results presented in Table \ref{windowsize}, it is evident that the window size of 300 outperforms both 100 and 200 in the TCN model.
Although the optimal results within the LSTM framework are achieved when the window size is 200, the $R^2$ and RMSE values for a window size of 300 are close to those obtained with the 200 window size configuration.
Consequently, in subsequent experiments, all LSTM models are implemented using a window size of 200, whereas TCN models utilize a window size of 300 as the input.

\begin{table}[]
\caption{Comparison of the performances of different window sizes in our experiment. The optimal results are highlighted in bold.}
\centering
\label{windowsize}
\begin{tabular}{ccccc}
\hline
\hline
\textbf{Model} & \textbf{Window Size} & \textbf{RMSE $\downarrow$} & \textbf{$R^2$ $\uparrow$}  \\
\hline
LSTM           & 100                  & 0.0177        & 0.9956      \\
LSTM           & 200                  & \textbf{0.0138}        & \textbf{0.9972}    \\
LSTM           & 300                  & 0.0144        & 0.9971            \\
\hline
TCN            & 100                  & 0.0306        & 0.9879       \\
TCN            & 200                  & 0.0375        & 0.9823       \\
TCN            & 300                  & \textbf{0.0288}        & \textbf{0.9894}   \\
\hline
\hline
\end{tabular}
\end{table}

Additionally, we perform feature augmentation on the seven input features.
The data augmentation in time series analysis is discussed by \cite{wen2020time}.
They highlight the complexity of augmenting multivariate time series data, particularly due to the need to consider potential complex dynamics across different features over time.
In order to preserve the interrelationships among variables in the original complex system, we compute the first-order and second-order derivatives separately for each of the original seven features and subsequently merge these derivative results with the original features.
This augmentation strategy allows us to expand the original feature set from 7 (1 time-independent feature + 6 time-dependent features) to either 13 or 19 while retaining the information of the original data.

We use the same experimental configuration of Table \ref{windowsize}, and compare the performance of different feature augmentation techniques.
The results of the experiment are shown in Table \ref{tab.augmentation}. 
It can be observed that while the impact of the same feature augmentation differs between the two models, the use of feature augmentation consistently leads to improved performance.
Compared to experiments conducted without feature augmentation, the second-order derivative feature augmentation resulted in a significant improvement of performance.

\begin{table}[]
\caption{Comparison of the performances of different feature augmentation techniques in our experiment. 
The 1$^{st}$-order derivatives represent new features derived from the gradients of the original features, while the 2$^{nd}$-order derivatives are derived from the 1$^{st}$-order derivatives.}
\centering
\label{tab.augmentation}
\begin{tabular}{ccccc}
\hline
\hline
\textbf{Model} & \textbf{Augmentation} & \textbf{RMSE $\downarrow$} & \textbf{$R^2$ $\uparrow$} \\
\hline
LSTM           & -                            & 0.0292        & 0.9885     \\
LSTM           & 1$^{st}$-order derivative       & 0.0171        & 0.9960     \\
LSTM           & 2$^{nd}$-order derivative      & \textbf{0.0138}        & \textbf{0.9972} \\
\hline
TCN            & -                            & 0.0475        & 0.9716      \\
TCN            & 1$^{st}$-order derivative       & 0.0338        & 0.9855      \\
TCN            & 2$^{nd}$-order derivative      & \textbf{0.0288}        & \textbf{0.9894} \\
\hline
\hline
\end{tabular}
\end{table}

\subsection{\label{sec34} Experiment Details}
In this subsection, details of all the experiments are presented, including the performance metrics, the training environment,the training optimizer selection and so on.
The root mean squared error (RMSE), the mean absolute scaled error (MASE) and the mean squared error (MSE) are the most common evaluation methods in time series forecasting tasks.
In addition, some evaluation metrics for regression tasks are used to evaluate our results more comprehensively, such as the coefficient of determination ($R^2$ or R-Squared) and the mean absolute error (MAE).
Physically, we used phase space distance (PSD) to measure the distance between the predicted time series and the real data.

RMSE and MSE are sensitive to large errors, in contrast to MAE, which is not sensitive to outliers, making MAE reflective of the overall predictive effectiveness of the models.
And MASE is useful for comparing the prediction accuracy for different data scales.
$R^2$ is a statistical measure that represents the proportion of variance for a dependent variable that's explained by an independent variable or variables in a regression model.
It is between 0 and 1, and the higher its value, the better the model fit.
In physics, PSD measures the difference between two points in phase space, considering all position and momentum coordinates.
It can give a comprehensive view of the differences in the system.
For all evaluation criteria, with the exception of $R^2$, a smaller value indicates better performance.

RMSE, MSE, MAE and $R^2$ can be expressed as Equation \ref{eq_ev_1}, \ref{eq_ev_2}, \ref{eq_ev_3} and \ref{eq_ev_4}, respectively:

\begin{eqnarray}
\label{eq_ev_1}
\text{RMSE} & = & \sqrt{\frac{1}{n} \sum_{i=1}^{n} (y_i - \hat{y}_i)^2},
\end{eqnarray}

\begin{eqnarray}
\label{eq_ev_2}
\text{MSE} & = & \frac{1}{n} \sum_{i=1}^{n} (y_i - \hat{y}_i)^2,
\end{eqnarray}

\begin{eqnarray}
\label{eq_ev_3}
\text{MAE} & = & \frac{1}{n}\sum_{i=1}^{n} |y_i - \hat{y}_i|,
\end{eqnarray}

\begin{eqnarray}
\label{eq_ev_4}
\text{$R^2$} & = & 1 - \frac{\sum_{i=1}^{n}(y_i - \hat{y}_i)^2}{\sum_{i=1}^{n}(y_i - \bar{y})^2},
\end{eqnarray}

where $n$ is the number of observations, $y_i$ represents the actual values, $\bar{y}$ denotes the mean of actual values, and $\hat{y}_i$ represents the predicted values. 
The formula of MASE is given below:
\begin{eqnarray}
\text{MASE} & = & \frac{\frac{1}{n}\sum_{i=1}^{n} |F_i - A_i|}{\frac{1}{n-1}\sum_{i=2}^{n} |A_i - A_{i-1}|},
\end{eqnarray}

where $n$ is again the number of observations, $F_i$ represents the predicted values and $A_i$ represents the actual values. 

In three-dimensional space, consider two points A and B.
$(q_{1A}, q_{2A}, q_{3A})$ is the position coordinate of point A and $(p_{1A}, p_{2A}, p_{3A})$ is the momentum coordinate of point A.
Point B is analogously described, with its position coordinates $(q_{1B}, q_{2B}, q_{3B})$ and momentum coordinates  $(p_{1B}, p_{2B}, p_{3B})$.
Then, PSD can be computed as:

\begin{eqnarray}
\text{PSD} & = & \sqrt{\sum_{i=i}^{n} (q_{iA} - q_{iB})^2 + (p_{iA} - p_{iB})^2}.
\end{eqnarray}

Next, the detailed of our experimental training setup will be presented.
Thanks to the release of Keras 3 \citep{chollet2015keras}, our deep models are developed in a mix of Keras 3 and Pytorch \citep{paszke2019pytorch}.
The model of LSTM is achieved by setting parameters and calling Pytorch's library, while the model of TCN comes from the work of \cite{lea2017temporal, KerasTCN}.
All of our experiments are run on a linux GPU server with an Intel i9-12900k 16-core CPU, a NVIDIA RTX-4090 GPU and 4 $\times$ 16 GB DDR4 memories.

In order to achieve a fair comparison, the random seed is set to a fixed number for all experiments.
In Section \ref{sec31}, the results demonstrate that a 3-layer LSTM with 512 hidden units outperforms other configurations in terms of performance.
Furthermore, in each convolutional layer with 256 filters and a kernel size equal to 5, the TCN network achieves the best performance relative to other architectures.
As per the experiments outlined in Section \ref{sec32}, the findings indicate a slight enhancement in performance when incorporating constraints on Hamiltonian conservation and angular momentum conservation within the loss function.
Additionally, the experimental results in Section \ref{sec33} illustrate that feature augmentation of input data effectively improve the prediction performance of both LSTM and TCN model.
It also demonstrates that LSTM and TCN models achieve optimal performance when the window size is set to 200 and 300, respectively.
Based on the aforementioned experiments, we employ a benchmark consisting of a 3-layer LSTM with 512 hidden units and a TCN with 256 filters and a kernel size of 5.
Moreover, our benchmark model incorporates feature augmentation with the second-order derivative feature augmentation method, and includes constraints on Hamiltonian conservation and angular momentum conservation within the loss function.

In the training process, the early stopping method \citep{prechelt2002early} is utilized to prevent overfitting.
The training will terminate prematurely when there is no improvement in validation performance for a specified number of iterations.
Additionally, in order to avoid the model from getting into a local optimum, we employ the dynamic learning rate adjustment strategy.
When the model does not show improvement on the validation set for a specified number of iterations, the learning rate is reduced by a factor of 10 from $1\times10^{-3}$ until it reached $1\times10^{-7}$.
We selected the Adam optimizer \citep{kingma2014adam} for training.
Given our utilization of the dynamic learning rate adjustment strategy, the learning rate for Adam is also set to dynamically adapt.
Lastly, all models are trained with floating-point operations in 64-bit precision.

\section{\label{sec4} Experimental Results}

\begin{table*}[htp]
\caption{Performance comparison of the optimal LSTM and TCN models for multivariate short sequence prediction.}
\label{tab:optimal_network}
\centering
\setlength{\tabcolsep}{4mm}{
\begin{tabular}{cccccccc}
\hline
\hline
\textbf{Model} & \textbf{Configuration} & \textbf{RMSE $\downarrow$} & \textbf{MSE $\downarrow$} & \textbf{MAE $\downarrow$} & \textbf{$R^2$ $\uparrow$} & \textbf{MASE $\downarrow$} & \textbf{PSD $\downarrow$} \\
\hline
LSTM & 3-layer, 512 hidden units & 0.0135 & 0.0002 & 0.0081 & 0.9974 & 0.0001 & 0.1318 \\
TCN & 5 kernel size, 256 filters & 0.0256 & 0.0007 & 0.0129 & 0.9919 & 0.0002 & 0.2356 \\
\hline
\hline
\end{tabular}
}
\end{table*}

\subsection{\label{sec41} Multi-Step Prediction}

\begin{figure}
\begin{center}
\includegraphics[scale=0.42]{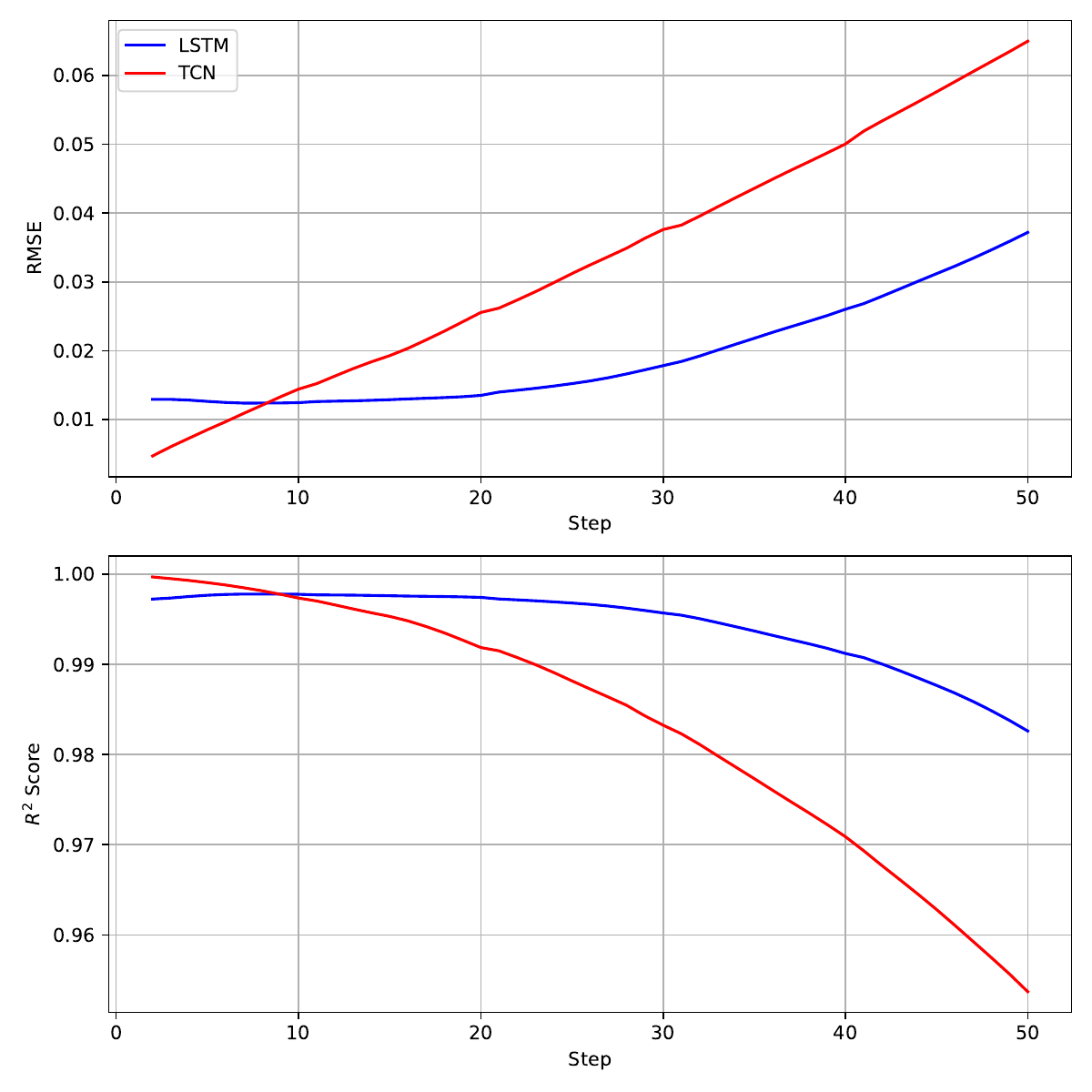}
\caption{Performance comparison of LSTM and TCN models across varying prediction counts on the test set. The Root Mean Square Error (RMSE), and the coefficient of determination ($R^2$) are shown from top to bottom respectively.}
\label{fig.comparison_plot}
\end{center}
\end{figure}

The results from Section \ref{sec3}, as displayed in Table \ref{tab:optimal_network}, show the superior performance of the optimal network architecture for 20-step predictions.
The LSTM outperforms the TCN across various evaluation metrics in this multivariate short-sequence prediction task.
Notably, both the LSTM and TCN achieved an $R^2$ score exceeding 0.99, underscoring their exceptional ability to predict the future multi-step time evolution of compact binary systems accurately.
It is evident that the LSTM model demonstrates a marginally lower RMSE of 0.0135 compared to the TCN's 0.0256, and a minimal MSE of 0.0002, suggesting a tighter clustering of predictions around the true values.
However, it is the PSD value where we see a more pronounced divergence: the LSTM's 0.1318 contrasts with the TCN's 0.2356, indicating a lower phase space distance from LSTM.
This comprehensive performance metric analysis confirms that while both models exhibit a strong ability to forecast the time evolution of compact binary systems, the LSTM holds a discernible edge in this multivariate short sequence prediction task.

Beyond the analysis at a fixed prediction length of 20 steps, this study further extends the comparison to encompass the performance of optimally configured LSTM and TCN models across a broader span of 1 to 50 prediction steps.
Figure \ref{fig.comparison_plot} delineates the progression of performance metrics (RMSE and $R^2$) for LSTM and TCN models on the test set as the prediction horizon extends from 1 to 50 steps.
As illustrated in Figure \ref{fig.comparison_plot}, within the various performance metrics, the TCN model consistently outperforms the LSTM within the initial 9 steps of prediction.
However, the TCN's performance begins to decline at a more accelerated rate than that of the LSTM from the 10th step.
As evidenced by the steeper slope of its decay curve, indicating the LSTM's enhanced robustness in longer step predictions.
This indicates that the TCN model is better suited for short-step sequence predictions, whereas the LSTM model demonstrates strengths for long-step sequence predictions.
Specifically, the RMSE of the LSTM model ascends more gradually with increasing prediction steps, whereas the TCN model indicates a steeper incline.
Both models experience a slight $R^2$ value decline as the number of prediction steps grows.
However, the LSTM's decrease is modest, maintaining a higher level of consistency.

\begin{figure}
\begin{center}
\includegraphics[scale=0.31]{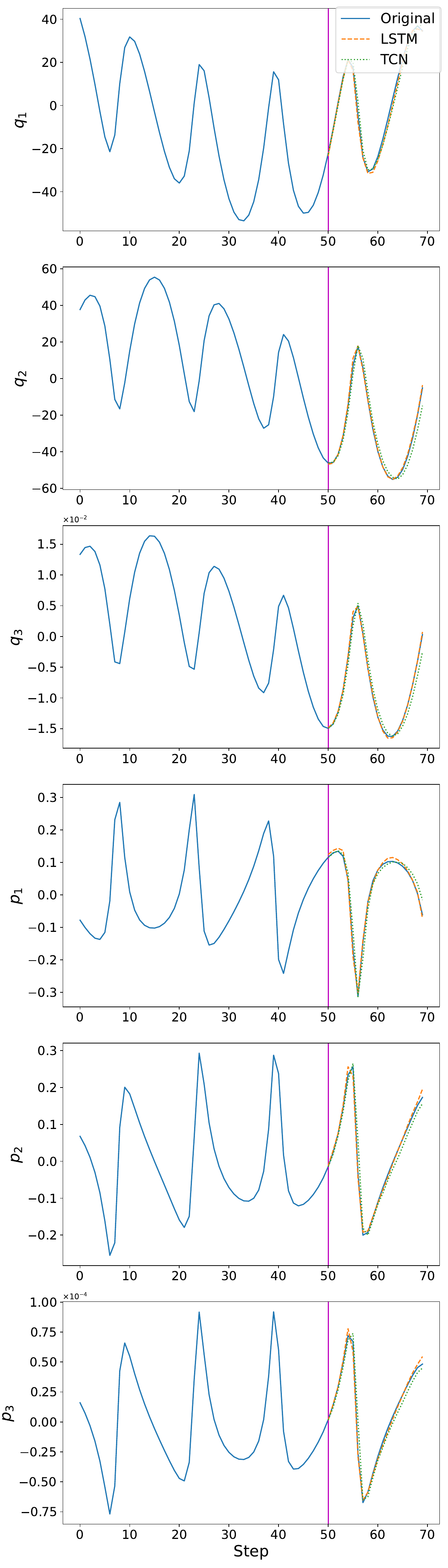}
\caption{Illustration of the comparison between the optimal LSTM and TCN 20 time steps predicted results. The evolution of the binary system is predicted with the original $C_4$ algorithm from 0 to 50 steps, as shown by the solid purple line.}
\label{fig.comparison_each_value_plot}
\end{center}
\end{figure}

Figure \ref{fig.comparison_each_value_plot} provides an intuitive comparison of the predicted results between the optimally configured LSTM and TCN models for 20 steps.
The first to third rows correspond to the three components of position coordinates, while the three components of momentum coordinates are shown in rows four to six.
The original sequence is represented by the blue solid line, with the LSTM and TCN predictions depicted by the orange and green dashed lines, respectively.
It is evident from the figure that both LSTM and TCN models closely fit the true values across all components, demonstrating significant robustness when forecasting sequences generated from various initial conditions.
However, despite the LSTM achieving commendable performance results, it consistently exhibits minor deviations in predicting peak values, as particularly observed in the sixth row of images in Figure \ref{fig.comparison_each_value_plot}, where the LSTM predictions slightly overshoot at the peaks.
In contrast, the TCN manages these peak situations with a nearly perfect adherence to the true values. 
This observation suggests that while TCN may not perform as well as LSTM in long-step sequence prediction, its precision in capturing sequence nuances holds promise for further performance enhancement through network improvements in future efforts.

\subsection{\label{sec42} Predicting Efficiency}

The implicit midpoint method is a well-established numerical technique employed in the solution of ordinary differential equations (ODEs) and the numerical integration of continuous functions.
It is characterized by its second-order accuracy and desirable stability properties, particularly in situations where explicit methods might suffer from stability constraints or stringent time-step restrictions.
However, implicit algorithms consume a lot of computational resources due to the need for multiple iterations, which can be reduced by optimizing the program to add multi-threaded parallel computation.
Additionally, there is a risk of convergence problems leading to the unavailability of numerical solutions.

In this study, not only is the implicit midpoint method employed as a benchmark for speed comparison, but we have also developed the extended phase-space (EPS) method.
To ensure a fair comparison, both the implicit midpoint method and the extended phase space method programming approach are used to predict the initial 30,000 steps of the time series, followed by predictions for the subsequent 2,000 steps using LSTM and TCN, respectively, to compare their inference times.
We compared the total time required for predicting the subsequent 2,000 steps using six different combinatory approaches.
These include the implicit midpoint method alone (Midpoint), the extended phase space method alone (EPS), Midpoint+LSTM, Midpoint+TCN, EPS+LSTM, and EPS+TCN.
The results are presented in six rows in Table \ref{tab.time_comparison}, each representing the predictive time required by these respective methods.
It is important to note that uniform downsampling is required before feeding the data into LSTM and TCN.
Specifically, this means that the original time series of 30,000 steps are reduced to 300 data points.
Similarly, for the original input sequence, the 20 steps predicted by LSTM and TCN are equivalent to 2,000 steps in the original sequence.
This approach is adopted not only because an excessive number of data points might degrade the performance of the model \citep{coolen2017replica}, but also because we have found that the data post-downsampling is sufficiently representative of the sequence information.

\begin{table}[]
\renewcommand{\arraystretch}{1.4}
\caption{Comparison of the time step efficiency for different combination methods, including the implicit midpoint method alone (Midpoint), the extended phase space method alone (EPS), Mid-point+LSTM, Mid-point+TCN, EPS+LSTM, and EPS+TCN.}
\centering
\label{tab.time_comparison}
\begin{tabular}{ccc}
\hline
\hline
\textbf{Combination} & \textbf{30000 Steps (s)} & \textbf{2000 Steps (s)}\\
\hline
Midpoint & $60.99\pm^{0.29}_{0.23}$ & $4.10\pm^{0.14}_{0.06}$ \\
Midpoint + LSTM & $60.98\pm^{0.16}_{0.16}$  & $\bm{0.01}\pm^{1.4\times10^{-4}}_{6.0\times10^{-5}}$ \\
Midpoint + TCN & $60.97\pm^{0.25}_{0.41}$ & $0.14\pm^{2.8\times10^{-3}}_{2.5\times10^{-3}}$ \\
\hline
EPS & $26.00\pm^{0.44}_{0.22}$ & $1.73\pm^{0.04}_{0.02}$ \\
EPS + LSTM & $25.98\pm^{0.30}_{0.15}$ & $\bm{0.04}\pm^{0.10}_{0.03}$ \\
EPS + TCN & $25.93\pm^{0.15}_{0.16}$ & $0.19\pm^{0.14}_{0.04}$ \\
\hline
\hline
\end{tabular}
\end{table}

All comparative experiments in this study are conducted on a CPU.
It is noteworthy that the prediction speed is expected to improve significantly if these deep learning models are to be executed on a GPU.
We conducted five replicate experiments for each method combination, calculating their average performance and variation range, to mitigate the impact of other uncontrolled factors.
The results demonstrate that LSTM is approximately 5 times more efficient than TCN in predicting 2,000 steps, and at least 40 times faster than the implicit midpoint method and the EPS method.
This underscores the substantial efficiency gains achieved by using deep learning models for modeling the time evolution of compact binary systems.
To highlight our contributions and innovative applications, we have deployed the inference functionality and time comparison experiments on the Hugging Face platform (\url{https://huggingface.co/spaces/EasonYan/MTECBS}), facilitating easier replication and validation of our research findings.

\subsection{\label{sec43} Simulation of gravitational waveform}

\begin{figure*}
\begin{center}
\includegraphics[scale=0.66]{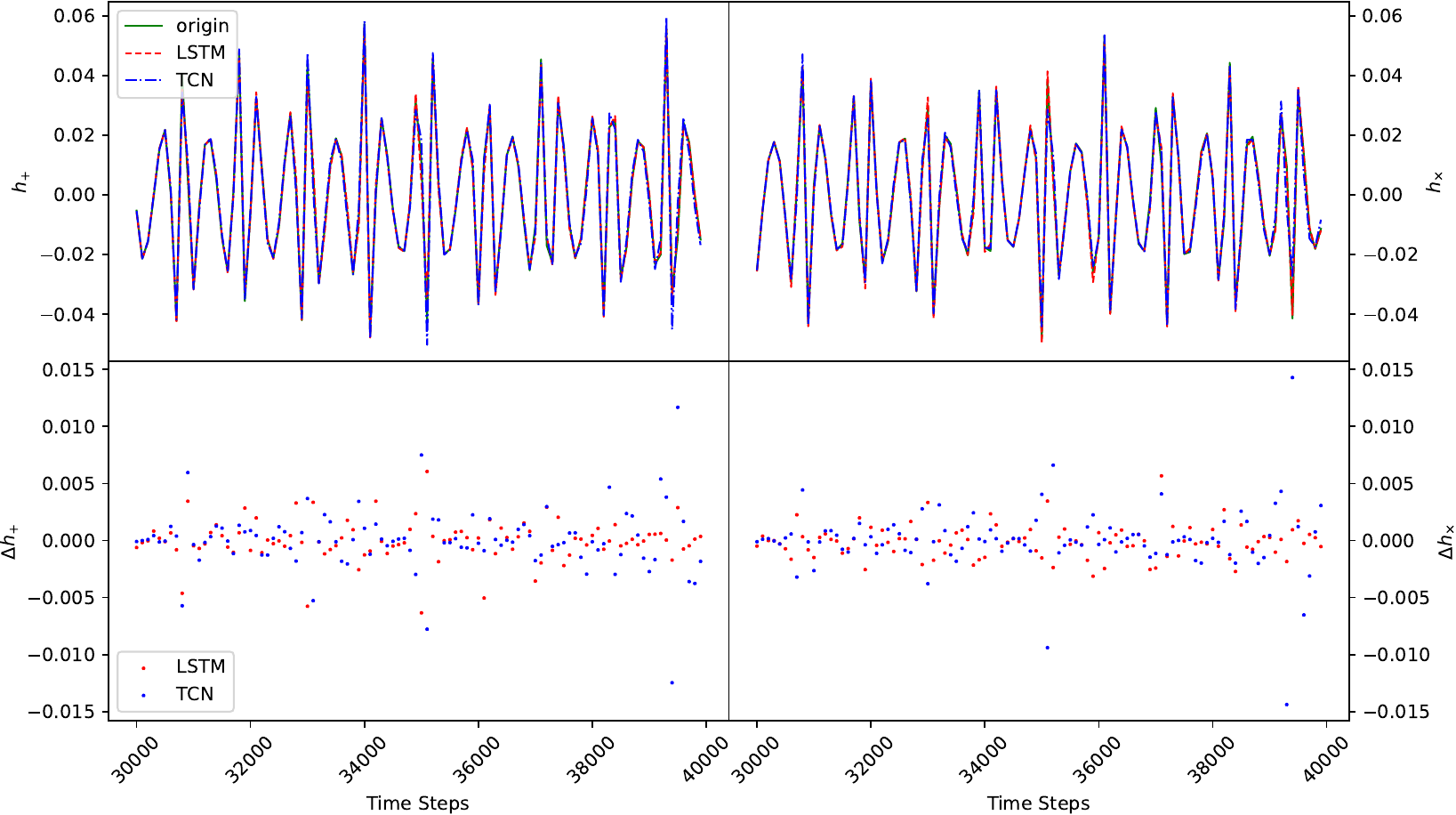}
\caption{Top left: the time evolution diagram of the plus-polarization component of the gravitational waveform $h_+$ drawn by $C_4$, LSTM, TCN. Top right: that of the cross-polarization component $h_{\times}$. Bottom left: LSTM and TCN deviate from the $h_+$ value calculated by $C_4$. Bottom right: similar plot as bottom left but for $h_{\times}$.}
\label{fig.h+andhx}
\end{center}
\end{figure*}

To rigorously assess the precision of LSTM and TCN models in simulating physical processes, we deliberately chose a specific configuration: setting the direction $\widehat{\mathbf{p}} = (1, 0, 0)$ and the orientation of the observer $\widehat{\mathbf{N}} = (0, \sin(\pi/4), \cos(\pi/4))$ with the following initial orbit conditions $(\beta;\textbf{r},\textbf{p})=(5; 35, 2, 0.1, 0.01, 0.15, 0.03)$. Furthermore, we fixed the distance between the observer and the source at $D = 1$, thereby establishing a controlled environment for comparative evaluation.

Employing the solutions of numerical algorithm $C_4$ as the benchmark, alongside the LSTM and TCN models, we generated time-evolution graphs for the cross-polarization ($h_{\times}$) and the plus-polarization ($h_+$) components of the gravitational waveforms. These visualizations strikingly revealed a high degree of overlap among the waveforms produced by all three methodologies, underscoring the remarkable accuracy with which both the LSTM and TCN models capture the underlying physics. Although minor deviations from the $C_4$ reference values were discernible, they were negligible in the context of the overall waveform representation.

To quantify these discrepancies, we plotted the discrete differences $\triangle h_{\times}$ and $\triangle h_+$, where $\triangle h_{\times}$ and $\triangle h_+$ denote the respective differences between the machine learning-predicted values ($h_{\times}^{ML}$, $h_+^{ML}$) and the true $C_4$-derived values ($h_{\times}^{C_4}$, $h_+^{C_4}$), i.e., $\triangle h_{\times} = h_{\times}^{ML} - h_{\times}^{C_4}$ and $\triangle h_+ = h_+^{ML} - h_+^{C_4}$. These plots offer a precise quantification of the models' deviations from the exact solution, substantiating the qualitative observation of their overall accuracy while providing a detailed assessment of the models' performance at individual time steps. In summary, the close correspondence between the LSTM and TCN waveforms and $C_4$ benchmark, as evident from both graphical overlays and the discrete difference plots, attests to the available accuracy of these machine learning techniques in modeling gravitational wave phenomena.

\section{\label{sec5} Conclusion \& Discussion}

This work presents a machine learning approach to simulating the temporal evolution of compact binary systems which are inherently characterized as nonlinear dynamical systems.
The Hamiltonian formalism is employed to describe the dynamics of the system, with the expending phase-space algorithm utilized to generate a large dataset of orbital trajectories for various celestial bodies, serving as a dataset for models such as TCN and LSTM.
The performance of the proposed models is compared against the correction map algorithms found in related research \citep{Luo_2020, Luo_2022}.

The study is conducted on both the TCN and LSTM models, examining the incremental inclusion of constraints and their impacts on performance metrics RMSE, $R^2$, and MSE.
The experimental results demonstrate that the incorporation of all physical constraints led to performance enhancements in both TCN and LSTM models.
For TCN, the results demonstrated that including physical conditions in the loss function effectively compensated for the model's fitting deficiencies.
We also find that using the Hamiltonian conservation and the angular momentum conservation alone leads to performance degradation.
This finding concurs with the conclusion presented in our previous work \citep{Luo_2022}.
In contrast, incorporating canonical equations into the loss function contributed to an improvement in model performance.

This work also evaluates the performance metrics of both LSTM and TCN models with different architectural configurations, including the number of layers, kernel size, the number of filters, and window size.
The results demonstrate that TCNs exhibit superior accuracy in short-term predictions, whereas LSTMs show better performance in long-term evolutionary predictions, effectively mitigating error growth.
Moreover, we have discovered that, owing to the advantages of an end-to-end model architecture, LSTM significantly outperforms TCN and other methods calculating first-order post-Newtonian approximations in terms of inference efficiency.

Our experiments demonstrate that both LSTM and TCN achieved an $R^2$ value exceeding 0.99 in multivariate short sequence prediction tasks, indicating their exceptionally high prediction accuracy.
Furthermore, we observed that TCN exhibits superior performance in predicting short-step sequences, whereas LSTM is better at predicting long-step sequences.
In terms of efficiency, LSTM is approximately 5 times more efficient than TCN in predicting 2,000 steps and is at least 40 times faster than numerical simulation methods such as the implicit midpoint method.
To highlight our research contributions and novelty, we have deployed the inference functionality and time comparison web service to an open-source website, thereby facilitating more convenient replication and validation of our research findings.

Conversely, LSTM and TCN models still face challenges in long sequence prediction.
In recent years, more advanced methods for time series prediction, such as the Transformer \citep{vaswani2017attention}, TimesNet \citep{wu2022timesnet}, and TSMixer \citep{chen2023tsmixer}, have been proposed.
These approaches have provided fresh insights for our future work.
Future research directions include exploring the synergistic predictive capabilities of TCN and LSTM models when used concurrently, and incorporating models like Transformer, TimesNet, TSMixer for comparison to assess their effectiveness in astrodynamical systems.

On the other hand, using PINNs for solving PDEs has gradually become mainstream.
In \cite{hu2020neural}, a method named Neural-PDE is proposed to use LSTM networks to learn and predict the numerical solutions of PDEs.
In \cite{mattheakis2022hamiltonian}, a Hamiltonian neural network is proposed as an equation-driven machine learning method to solve the differential equations in the govern dynamical systems.
\cite{greydanus2019hamiltonian} presents a novel neural network model, inspired by Hamiltonian mechanics, which learns to follow exact conservation laws in an unsupervised manner by particularly addressing the differential equations that govern the dynamics of physical systems.
Our future work will aim to focus on an optimal combination of sequence predictions without much prior knowledge or inductive bias from physics models and solutions from PINNs given our physical model with the Hamiltonian for compact binary systems as there are advantages and disadvantages with both approaches.

Additionally, efforts will be directed towards optimizing data sampling schemes to enhance the efficiency of machine learning processes in this context.
Moreover, we will incorporate a gravitational radiation term to compute the actual gravitational waveforms.
In addition to employing machine learning techniques to grasp the evolutionary patterns of the binary without the dissipated term, we will attempt to establish mappings between the gravitational waveforms and various parameters of the binary system, such as the masses, periods, direction of the wave source, and distance.
With the knowledge of the waveform, these parameters can be directly inferred, obviating the need for manual application of matched filtering algorithms to analyze gravitational wave data, which typically requires searching for the most prominent signal superposition to deduce the characteristics of the binary.

%
%
%
%

\begin{acknowledgments}
This work is supported in part by the National Natural Science Foundation of China (NSFC) under Grant Nos. 12203108, 11875327, 12275367 and 12073089, the Fundamental Research Funds for the Central Universities, and the Sun Yat-sen University Science Foundation.
\end{acknowledgments}

\bibliography{sample631}{}
\bibliographystyle{aasjournal}



\end{document}